\documentclass[twocolumn,english,superscriptaddress,prl,amsmath,amssymb,floatfix,longbibliography]{revtex4-2}
\usepackage[T1]{fontenc}
\usepackage[utf8]{inputenc}
\usepackage{color}
\usepackage{babel}
\usepackage{bm}
\usepackage{amsmath}
\usepackage{graphicx}
\usepackage{esint}
\usepackage[unicode=true,pdfusetitle,
 bookmarks=true,bookmarksnumbered=false,bookmarksopen=false,
 breaklinks=false,pdfborder={0 0 1},backref=false,colorlinks=true]
 {hyperref}
\hypersetup{
 pdfborderstyle=}

\makeatletter
\usepackage{babel}

\usepackage{dcolumn}
\usepackage{bm}
\usepackage{color}
\usepackage{soul}
\usepackage{amsfonts}
\usepackage{slashed}
\usepackage{enumerate}

\makeatother

\begin{document}
\global\long\def\sgn{\mathrm{sgn}}%
\global\long\def\ket#1{\left|#1\right\rangle }%
\global\long\def\bra#1{\left\langle #1\right|}%
\global\long\def\sp#1#2{\langle#1|#2\rangle}%
\global\long\def\abs#1{\left|#1\right|}%
\global\long\def\avg#1{\langle#1\rangle}%

\title{Extracting the scaling dimension of quantum Hall quasiparticles from
current correlations}
\author{Noam Schiller}
\affiliation{Department of Condensed Matter Physics, Weizmann Institute of Science,
Rehovot, 76100 Israel}
\author{Yuval Oreg}
\affiliation{Department of Condensed Matter Physics, Weizmann Institute of Science,
Rehovot, 76100 Israel}
\author{Kyrylo Snizhko}
\affiliation{Institute for Quantum Materials and Technologies, Karlsruhe Institute
of Technology, 76021 Karlsruhe, Germany}
\affiliation{Department of Condensed Matter Physics, Weizmann Institute of Science,
Rehovot, 76100 Israel}
\affiliation{Univ. Grenoble Alpes, CEA, Grenoble INP, IRIG, PHELIQS, 38000 Grenoble, France}
\begin{abstract}
Fractional quantum Hall quasiparticles are generally characterized
by two quantum numbers: electric charge $Q$ and scaling dimension
$\Delta$. For the simplest states (such as the Laughlin series), the
scaling dimension determines the quasiparticle's anyonic statistics
(the statistical phase $\theta=2\pi\Delta$). For more complicated
states (featuring counterpropagating modes or non-Abelian statistics),
knowing the scaling dimension is not enough to extract the quasiparticle
statistics. Nevertheless, even in those cases, knowing the scaling
dimension facilitates distinguishing different candidate theories
for describing the quantum Hall state at a particular filling (such
as PH-Pfaffian and anti-Pfaffian at $\nu=5/2$). Here, we propose a
scheme for extracting the scaling dimension of quantum Hall quasiparticles
from thermal tunneling noise produced at a quantum point contact.
Our scheme makes only minimal assumptions about the edge structure and features the level of robustness, simplicity, and model independence comparable
to extracting the quasiparticle charge from tunneling shot noise.
\end{abstract}
\maketitle

\section{\label{sec:I_Introduction}Introduction}

The fractional quantum Hall (FQH) effect is renowned as a showcase
example of strongly correlated quantum states. Electron-electron interactions
result in the emergence of fractional quasiparticles that are predicted
to possess fractional charge and fractional statistics \citep{QHE_Book,Stormer1999,Girvin2005,Arovas1984,WenCLL,FrohlichAnomaly,Wen1991a,WenReview}.
For some filling factors, the fractional statistics is expected to
be non-Abelian, which can be instrumental for topologically protected
quantum computation \citep{Nayak2008}.

\begin{figure}
\begin{centering}
\includegraphics[width=1\columnwidth]{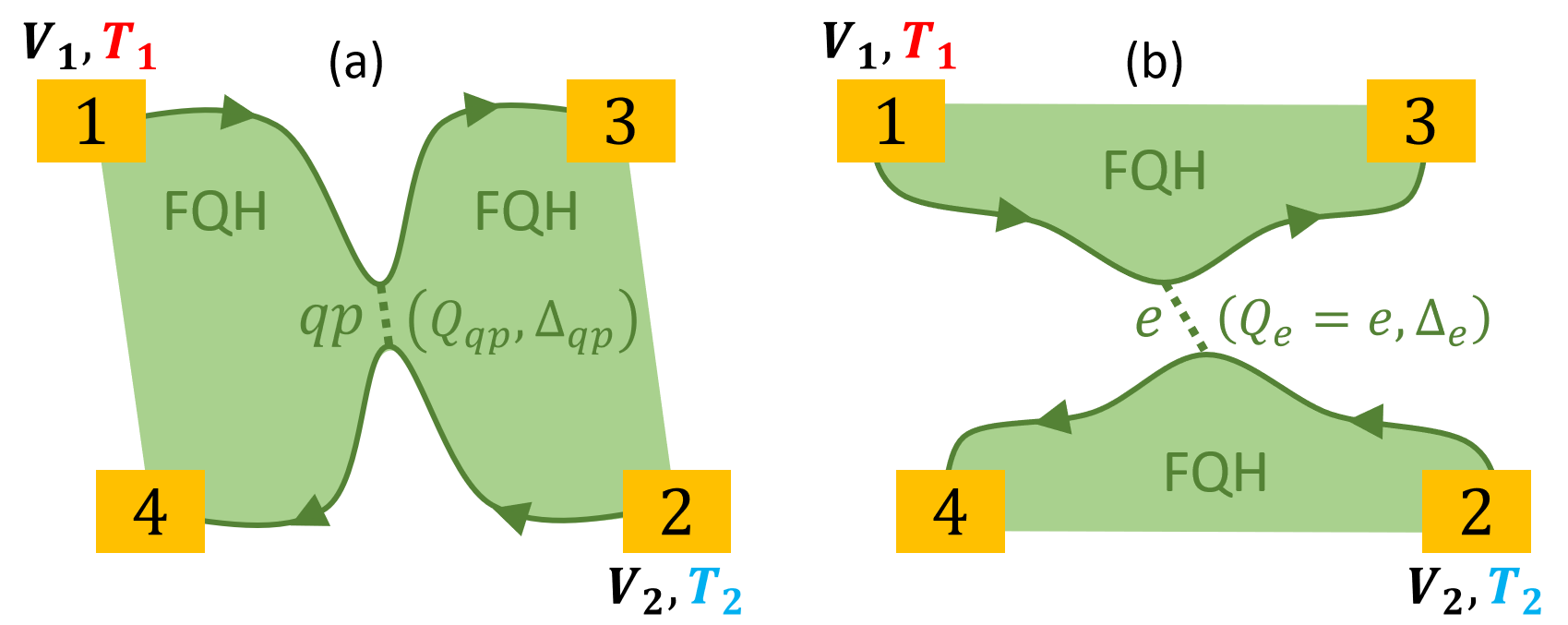}
\par\end{centering}
\caption{\label{fig:QPC_setup}The standard setups for investigating quasiparticle
{[}$qp$, panel (a){]} and electron {[}$e$, panel (b){]} tunneling
in quantum Hall systems. Ohmic contacts 1 and 2 having voltages $V_{1}$
and $V_{2}$ respectively are used to inject electric current to the
edges. Two FQH edges meet in the middle of the device, giving rise
to the tunneling current $I_{T}$ and noise $S$, which are inferred
from measurements at contacts 3 and 4. The dependence of $S$ and
$I_{T}$ on $V_{1}-V_{2}$ enables a reliable extraction of the charge
of the tunneling particle ($qp$ or $e$). Here we propose to extract
the tunneling particle's scaling dimension, $\Delta$, which governs
the edge dynamics, using the temperatures $T_{1}$ and $T_{2}$ of
the injected edge modes as additional experimental knobs. In the simple
Abelian cases the exchange statistics of the tunneling particle is
related to $\Delta$. }
\end{figure}

The fractional charge of FQH quasiparticles has numerous confirmations
obtained with a number of methods \citep{Goldman1995,FracChargeObs_Heiblum,FracChargeObs_Glattli,Griffiths2000,Chung2003,Martin2004,Dolev2008,Dolev2010,Venkatachalam2011,Mills2020,Roosli2021}.
The most used method for extracting the quasiparticle charge is based
on measuring the shot noise at a quantum point contact (QPC) where
two FQH edges meet and quasiparticle tunneling processes take place
\citep{FracChargeObs_Heiblum,FracChargeObs_Glattli,Griffiths2000,Chung2003,Dolev2008,Dolev2010},
cf.~Fig.~\ref{fig:QPC_setup}. At the same time, the first promising
measurements of the fractional statistics have been obtained only
recently \citep{Bartolomei2020,Nakamura2020a}, despite a large number
of distinct theoretical proposals \citep{Kane2003,Sarma2005,Stern2006,Bonderson2006,Bonderson2006a,Law2006,Ponomarenko2010,Campagnano2012,Rosenow2016,Han2016,Grass2020}.

With statistics measurements not readily available, an important problem
in the field is discriminating between different candidate theories
that can describe the same filling factor. For example, a number of
theories can describe $\nu=5/2$, some host non-Abelian quasiparticles,
while others do not \citep{Willett2013,Heiblum2020}. A basic approach
to discriminating between different theories relies on extracting
two key properties of the fundamental quasiparticle: its electric
charge $Q$ and its scaling dimension $\Delta$ \citep{Radu_experiment,BoCheFro_RaduExp_analysis,Ensslin_FQHE_TunnExp}.
The charge alone does not always allow one to discriminate different
theories. For example, in most candidate theories for $\nu=5/2$,
the fundamental quasiparticle has charge $Q=e/4$.

A series of theoretical works, based on gauge invariance and the topological
nature of the bulk states, have come to the following conclusions
\citep{Wen2004,Jain2007}: The scaling dimension of the
quasiparticle, closely related to the parameter $K$ of non-chiral Luttinger
liquids \citep{Giamarchi2003}, characterizes its dynamics at a FQH
edge. In the simplest cases, when only modes of a single chirality
are present on the edge, the scaling dimension is directly related
to the quasiparticle braiding statistics. The statistical phase is
then given by $\theta=2\pi\Delta$. In the case of non-Abelian statistics
of quasiparticles, the scaling dimension may only capture its Abelian
part. For more complicated edge structures featuring counterpropagating
modes \citep{KaneFisherPolchinski,KaneFisher,Wang2013a,Sun2020d},
the scaling dimension may not correspond to the quasiparticle statistics
at all. Nevertheless, even then the scaling dimension is an important
property characterizing the quasiparticle behavior and allowing one
to discriminate different candidate theories.

The so far most promising attempts to measure the scaling dimension
concerned the dependence of the tunneling current at a QPC on the
voltage bias between the two edges \citep{Radu_experiment,Ensslin_FQHE_TunnExp}.
The scheme should simultaneously extract both $Q$ and $\Delta$.
Some experiments produce data that are congruent with the theoretically
predicted curves \citep{Radu_experiment,Ensslin_FQHE_TunnExp}. However,
the extracted values of $Q$ and $\Delta$ vary greatly between different
experimental configurations. Furthermore, in none of the configurations
do both the fitted charge and the scaling dimension agree with those
predicted by expected candidate theories \citep{Ensslin_FQHE_TunnExp}.
Other experiments of the same type report that the tunneling current
dependence on voltage significantly deviates from theoretically predicted
curves \citep{Roddaro2003,Chung2003,Roddaro2005,Heiblum2006}. One
possible explanation for such behavior is that changes in the bias
voltage, $V=V_{1}-V_{2}$, cf.~Fig.~\ref{fig:QPC_setup}, affect
the electrostatic balance at the QPC, changing its shape and the tunneling
matrix element; the dependence of the tunneling amplitude on the voltage
in turn alters the behavior of the tunneling current in a non-universal
way.

This non-universality can be excluded by considering the ratio $F=S/(2eI_{T})$
(also called the Fano factor) of the excess noise $S$ appearing due
to tunneling (the noise measured at contact 3 in the setups of Fig.~\ref{fig:QPC_setup}
minus the Johnson-Nyquist noise $2\nu e^{2}k_{B}T_{1}/h$) 
to the tunneling current $I_{T}$. When the tunneling amplitude $\eta$
is small, both are $\propto\abs{\eta}^{2}$, so that the ratio $F$
does not depend on $\eta$. In fact, this is the very trick that enables
reliable determination of the quasiparticle charge in such setups
\citep{FracChargeObs_Heiblum,FracChargeObs_Glattli,Griffiths2000,Chung2003,Dolev2008,Dolev2010}.
It has been theoretically shown that considering the dependence of
$F$ on bias voltage at non-zero temperatures, in principle, allows
for extracting not only charge but also the scaling dimension \citep{Snizhko2015}.
However, the term involving the scaling dimension turns out to be
only a small correction to the main charge-dependent term and, therefore,
cannot be reliably extracted from the standard experiments.

Considering temperature dependence instead of the voltage dependence
is a new promising direction. On one hand, recent experiments developed
a way of changing the edge temperature in a quick and electrically
controllable manner \citep{Jezouin2013,Banerjee2017,Banerjee2018,Rosenblatt2020,Srivastav2021,Melcer2021,Kumar2021b}.
On the other hand, a number of theoretical works considered the QPC
physics when the two edges are at different temperatures \citep{Takei2011,ShtaSniChe_2014,Snizhko2016,Rech2020}.
In particular, an intriguing effect of the excess noise dropping when
the temperature imbalance between the edges is increased has been
predicted \citep{Rech2020}.

In the present paper, we study the dependence of the Fano factor $F$
at the QPC on the temperatures of the two edges. We show that the
scaling dimension can be extracted from the Fano factor's temperature
dependence. The paper is organized as follows. We briefly describe
our key results in Sec.~\ref{sec:II_Main_Results}. We then describe
in Sec.~\ref{sec:III_Noise_results} the model used to obtain these
results for the noise in the setups of Fig.~\ref{fig:QPC_setup}.
In Sec.~\ref{sec:IV_experimentally_relevant_scenarios} we analyze
our predictions for some experimentally relevant scenarios. A discussion of the factors that may render the scaling dimension non-universal and of other experimental subtleties that may restrict the applicability of the proposed method is provided in Sec.~\ref{sec:V_Discussion}. We conclude
with Sec.~\ref{sec:VI_Conclusion}. For completeness,
we provide a brief overview of the basics of the quantum Hall edge
theory and of the meaning of the quasiparticles' scaling dimensions
in Appendix~\ref{app:A_edge_theory_background}. Technical details
of the derivation of our results are relegated to Appendix~\ref{app:B_Derivations}.

\section{\label{sec:II_Main_Results}Main Results}

\begin{figure}
\begin{centering}
\includegraphics[width=1\columnwidth]{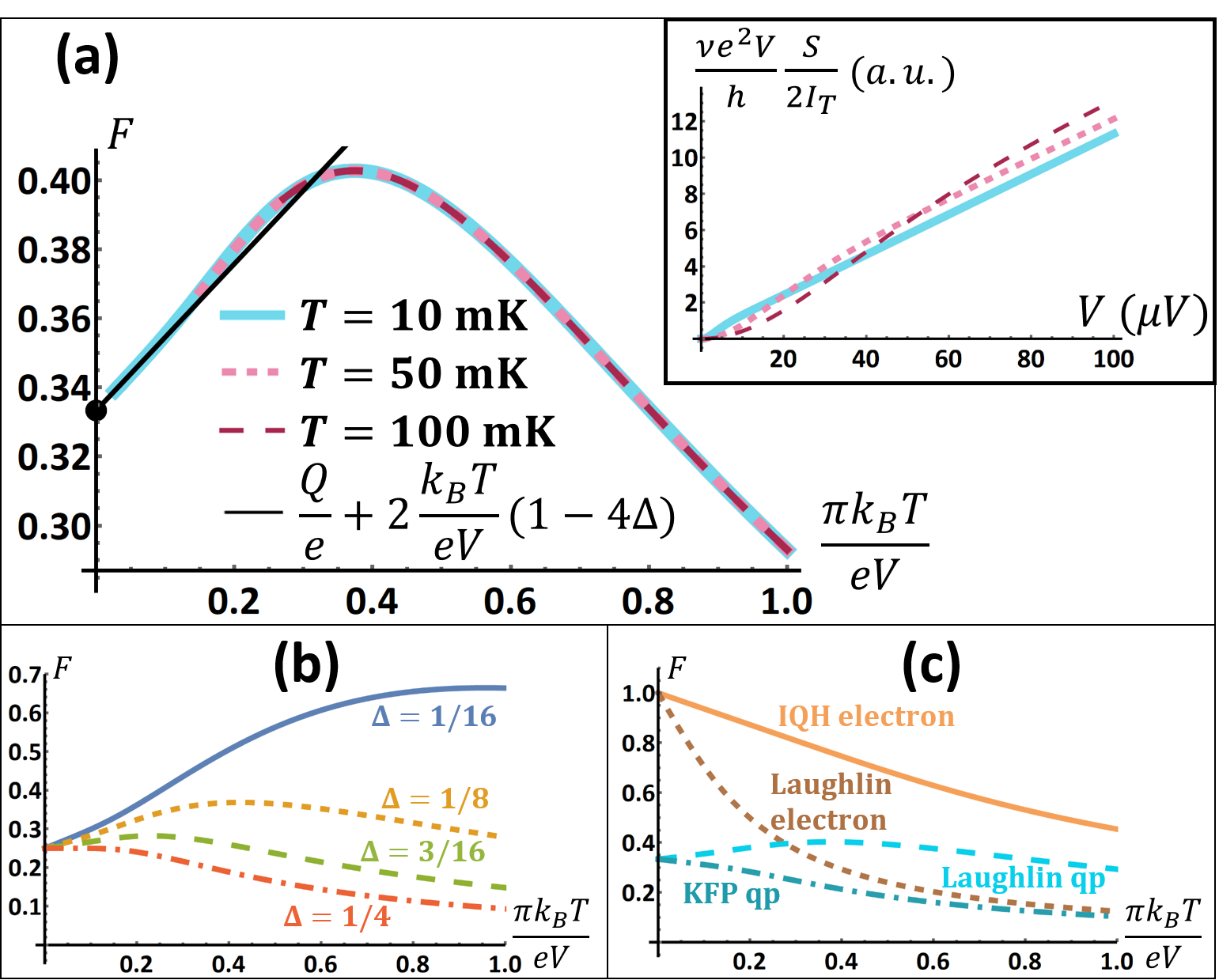}
\par\end{centering}
\caption{\label{fig:sameTemps_summary}Fano factor's dependence on temperature
$T=T_{1}=T_{2}$ (cf.~Fig.~\ref{fig:QPC_setup}) as a tool for extracting
the scaling dimensions of FQH quasiparticles and electrons. (a)~The
dependence of the Fano factor, $F=S/(2eI_{T})$ in Eq.~(\ref{eq:Fano_sameTemps}),
on $T/V$ for $\nu=1/3$ Laughlin quasiparticles $(Q=e/3,\Delta=1/6)$.
The curve for each temperature corresponds to $V\in[0,100]\,\mathrm{\mu V}$.
The curves collapse on top of each other, showing the universal behavior
of the Fano factor. The thin black line corresponds to the asymptotic
behavior in Eq.~(\ref{eq:Fano_sameTemps_asymp}). The dot on the
vertical axis corresponds to $F=Q/e$. Inset:~The same data plotted
in the way experiments are conventionally analyzed. The curves for
different temperatures appear unrelated. (b)~Fano factor vs. $T/V$
for quasiparticles that appear in various candidate theories of $\nu=5/2$
FQH effect: $Q=e/4$ (in all the theories), $\Delta=1/16$ ($K=8$ theory),
$1/8$ (Pfaffian and PH-Pfaffian), $3/16$ (331 state) or $1/4$ (anti-Pfaffian)
\citep{Yang2013}. (c)~Fano factor vs. $T/V$ for $\nu=1/3$ Laughlin
quasiparticle $(Q=e/3,\Delta=1/6)$, one of the quasiparticles in
the Kane-Fisher-Polchinski (KFP) $\nu=2/3$ fixed-point theory $(Q=e/3,\Delta=1/3)$,
non-interacting integer quantum Hall (IQH) electrons $(Q=e,\Delta=1/2)$,
electrons in the $\nu=1/3$ Laughlin state $(Q=e,\Delta=3/2)$ \citep{WenCLL,KaneFisherPolchinski,KaneFisher}.}
\end{figure}

We first consider the case of equal temperatures of the two edges:
$T_{1}=T_{2}=T$. In the inset of Fig.~\ref{fig:sameTemps_summary}(a),
we present the dependence of noise on the bias voltage for several
temperatures $T$. Notice that we plot $\nu e^{2}V/h\times S/(2I_{T})=\nu e^{2}V/h\times eF$
so that the dependence of the tunneling amplitude on the bias voltage
or the temperature cancels out. 
The slope at large $V$ corresponds to the charge of the tunneling
quasiparticle. However, otherwise the curves do not appear to exhibit
universality. In Appendix~\ref{app:Equal_Temperatures} we show that
in the regime of weak tunneling of quasiparticles or electrons across
the QPC (cf.~Fig.~\ref{fig:QPC_setup}), the Fano factor is a universal
function of $T/V$, namely
\begin{equation}
F=\frac{2Q}{\pi e}\mathrm{Im}\left[\psi\left(2\Delta+i\frac{QV}{2\pi k_{B}T}\right)\right],\label{eq:Fano_sameTemps}
\end{equation}
where $Q$ and $\Delta$ are respectively the charge and the scaling
dimension of the quasiparticle that tunnels, $k_{B}$ is the Boltzmann
constant, the digamma function $\psi(x)=\Gamma'(x)/\Gamma(x)$ is the logarithmic derivative
of the gamma function, and $\mathrm{Im}$ stands for the imaginary
part. This universality is illustrated in Fig.~\ref{fig:sameTemps_summary}(a)
where the curves of the inset of Fig.~\ref{fig:sameTemps_summary}(a)
are plotted as a function of $T/V$. Fitting experimental data to
this curve should enable reliable extraction of the charge and the
scaling dimension.

Further, consider the limit $eV\gg k_{B}T$, which corresponds to
a typical regime investigated experimentally. Then
\begin{equation}
\left.F\right|_{eV\gg k_{B}T}=\frac{Q}{e}+2\frac{k_{B}T}{eV}(1-4\Delta)+O\left[\left(\frac{k_{B}T}{eV}\right)^{2}\right].\label{eq:Fano_sameTemps_asymp}
\end{equation}
The first term of the expression represents the well-known result
that the shot-noise Fano factor corresponds to the charge of the tunneling
quasiparticle. The scaling dimension enters the second, subleading
term. This subleading correction is a linear function of $T/V$ and
can be quite sizeable, cf.~Fig.~\ref{fig:sameTemps_summary}(a).

In Fig.~\ref{fig:sameTemps_summary}(b,c), we give several examples
of the Fano factor behavior for quasiparticles corresponding to various
quantum Hall states. Notice that quasiparticles of the same charge
but different scaling dimensions produce notably different curves.
In addition, we emphasize that the strongly-interacting nature of
FQH states is manifest not only in the existence of fractional quasiparticles,
but also in the electron scaling dimension, which can be inferred
by the proposed method, cf.~Figs.~\ref{fig:sameTemps_summary}(c)
and \ref{fig:QPC_setup}(b).

The origin of the above correction term can be roughly related to
the quasiparticle exclusion statistics. Consider as an example non-interacting
edges of $\nu=1$ integer quantum Hall states. At $T=0$, each edge is a
Fermi sea of electrons occupied up to each edge's Fermi level. Only
the window of energies $eV$, where the electrons on one edge are
not balanced by the electrons of the other edge is important. The
electrons of different energies within this window tunnel independently,
so the Pauli exclusion principle does not affect the observable quantities.
At $T>0$, the edges of the Fermi seas become smeared. An electron
within the $eV$ window can be hindered from tunneling to the other
edge if this energy level is occupied (which happens with a $T$-dependent
probability). This reduces the fluctuations of the occupation of this
level, and thus reduces the noise and the Fano factor. This intuitive
picture agrees with the prediction of Eq.~(\ref{eq:Fano_sameTemps_asymp})
as for non-interacting electrons $\Delta=1/2$.

Had the electrons attracted each other or tended to bunch (as bosons
do), the presence of an electron at an energy level before the QPC
would increase the probability of tunneling of another electron from
the opposite edge and would increase the noise. This is the case,
e.g., when dealing with $\nu=1/3$ Laughlin quasiparticles (that can
bunch up to $3$ in one quantum state and have $\Delta=1/6$). It
is remarkable that quasiparticles with $\Delta=1/4$, (for example,
semions) which for sufficiently simple edges are half-way between
bosons and fermions in terms of the statistical phase $\theta=2\pi\Delta$,
would produce no correction to the Fano factor in this regime.

\begin{figure}
\begin{centering}
\includegraphics[width=1\columnwidth]{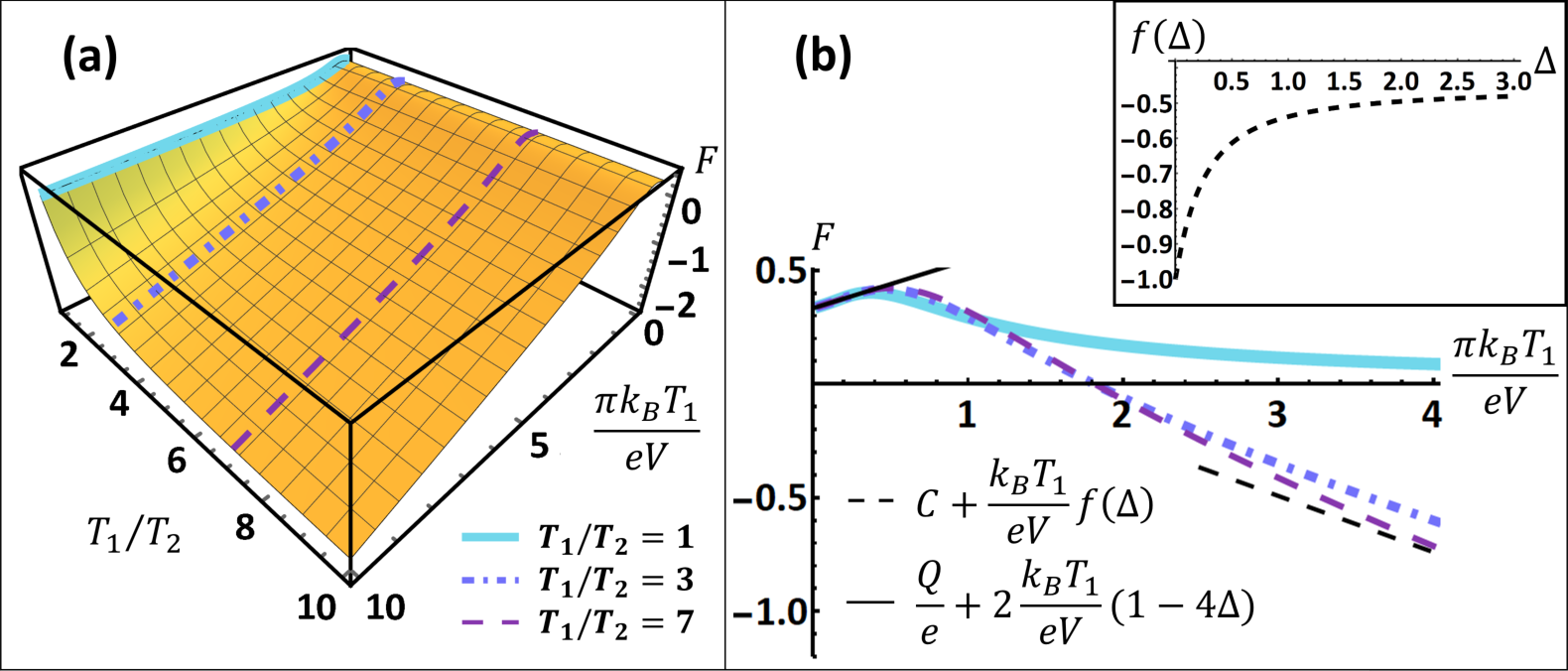}
\par\end{centering}
\caption{\label{fig:diffTemps_summary}The Fano factor for the case of $T_{1}\protect\geq T_{2}$
(cf.~Fig.~\ref{fig:QPC_setup}). (a)~Plot of the universal function
given in Eq.~(\ref{eq:Fano_diffTemps}) illustrated for the Laughlin
quasiparticle $(Q=e/3,\Delta=1/6)$. Lines denote the cuts corresponding
to $T_{1}/T_{2}=1$, $3$, and $7$. (b)~Cuts corresponding to $T_{1}/T_{2}=1$,
$3$, and $7$ plotted as a function of $T_{1}/V$. At small $T_{1}/V$
the Fano factor's behavior is described by Eq.~(\ref{eq:Fano_diffTemps_highV_asymp})
(solid black line). At large $T_{1}/V$, Eq.~(\ref{eq:Fano_diffTemps_highT1_asymp})
gives the asymptotic behavior (dashed black line); the slope of the
curves at $T_{1}/T_{2}=3$ and $7$ is quite close to the expected
behavior, while the offset $C$ only converges to zero when $T_{1}/T_{2}\rightarrow\infty$.
Inset---The slope $f(\Delta)$ entering Eq.~(\ref{eq:Fano_diffTemps_highT1_asymp})
as a function of the quasiparticle scaling dimension.}
\end{figure}

It is important to emphasize two things. First, while the above consideration
is qualitative and appeals to the intuition of non-interacting systems,
we have derived Eqs.~(\ref{eq:Fano_sameTemps}) and (\ref{eq:Fano_sameTemps_asymp})
using proper quantum Hall edge theory that takes the interacting nature
of the FQH states into account. Second, the relation between the
scaling dimension and quasiparticle statistics is not universal and
holds only for sufficiently simple Abelian quantum Hall edges, when
only modes of a single chirality are present on the edge. Therefore,
it is correct to characterize the noise in terms of the scaling dimension
$\Delta$, and not in terms of the quasiparticle statistics.

We next consider the situation of general temperatures, assuming without
loss of generality $T_{1}\geq T_{2}$. In this case, the Fano factor
can be expressed as a universal function of $T_{1}/V$ and $T_{1}/T_{2}$:
\begin{equation}
F=\mathcal{F}_{Q,\Delta}\left(\frac{\pi k_{B}T_{1}}{eV},\frac{T_{1}}{T_{2}}\right),\label{eq:Fano_diffTemps}
\end{equation}
where the detailed expression is presented in Sec.~\ref{sec:III_Noise_results},
Eq.~(\ref{eq:Fano_Integral}). The universality is manifest in the
function being determined only by the charge $Q$ and the scaling
dimension $\Delta$ of the tunneling quasiparticle. All data
for the Fano factor dependence on the bias voltage $V$ and the
two temperatures, $T_{1}$ and $T_{2}$, should collapse on a two-dimensional (2D)
surface, which is determined only by the values of $Q$ and $\Delta$.
We illustrate the behavior of this function for Laughlin quasiparticles
in Fig.~\ref{fig:diffTemps_summary}(a).

The expression in Eq.~(\ref{eq:Fano_diffTemps}) simplifies in some
limiting cases. First, we show in Appendix~\ref{app:Equal_Temperatures}
that for $T_{1}=T_{2}=T$, Eq.~(\ref{eq:Fano_diffTemps}) reduces
to Eq.~(\ref{eq:Fano_sameTemps}). Second, we show in Appendix~\ref{app:Dominant Voltage}
that in the regime $eV\gg k_{B}T_{1}\geq k_{B}T_{2}$, Eq.~(\ref{eq:Fano_diffTemps})
simplifies to\footnote{The absence of symmetry between $T_{1}$ and $T_{2}$ in this expression
is due to the fact that the noise is measured at contact 3 in Fig.~\ref{fig:QPC_setup},
which is located on the edge with temperature $T_{1}$. If the noise
is measured at contact 4, the roles of $T_{1}$ and $T_{2}$ are exchanged.
Note also the similarity of this expression to Eq.~(\ref{eq:Fano_sameTemps_asymp}).}
\begin{equation}
\left.F\right|_{eV\gg k_{B}T_{1,2}}=\frac{Q}{e}+2\frac{k_{B}T_{1}}{eV}(1-4\Delta)+O\left[\left(\frac{k_{B}T_{1,2}}{eV}\right)^{2}\right].\label{eq:Fano_diffTemps_highV_asymp}
\end{equation}
Third, we show in Appendix~\ref{app:Dominant_Temperature} that in
the regime dominated by the temperature imbalance, $k_{B}T_{1}\gg eV,\,k_{B}T_{2}$,
\begin{equation}
\left.F\right|_{k_{B}T_{1}\gg eV,\,k_{B}T_{2}}=\frac{k_{B}T_{1}}{eV}f(\Delta)+O\left[\frac{eV}{k_{B}T_{1}},\frac{k_{B}T_{1}}{eV}\left(\frac{T_{2}}{T_{1}}\right)^{2}\right],\label{eq:Fano_diffTemps_highT1_asymp}
\end{equation}
where $f(\Delta)$ is a function of the scaling dimension, whose behavior
is presented in the inset of Fig.~\ref{fig:diffTemps_summary}(b).
Notice that the bias voltage enters the expression of Eq.~(\ref{eq:Fano_diffTemps_highT1_asymp}),
but the quasiparticle charge $Q$ does not. We illustrate the asymptotic
behaviors of Eqs.~(\ref{eq:Fano_diffTemps_highV_asymp}, \ref{eq:Fano_diffTemps_highT1_asymp})
in Fig.~\ref{fig:diffTemps_summary}(b).

Note that the Fano factor can become negative when $T_{1}\neq T_{2}$,
cf.~Fig.~\ref{fig:diffTemps_summary}. This is due to the excess
noise becoming negative in the presence of temperature imbalance between
the edges, similarly to the effect predicted recently in Ref.~\citep{Rech2020}.
We emphasize the non-triviality of this result: while raising the
temperature of one edge is expected to affect the intensity of the
tunneling processes, the Fano factor's behavior shows that the noise
and the current are affected in different ways. The total noise at
contact $3$ (cf.~Fig.~\ref{fig:QPC_setup}) comprises the excess
noise $S$, as well as the Johnson--Nyquist noise of the FQH edge
(which is typically subtracted in experiments). The latter, of course,
grows as $T_{1}$ is increased, as does the total noise at contact
$3$. However, the excess noise $S$ involving the contribution of
the tunneling processes becomes negative.\footnote{Our prediction and the prediction of Ref.~\citep{Rech2020} are complementary
in a non-trivial way. First, Ref.~\citep{Rech2020} considered the
case of small differences between $T_{1}$ and $T_{2}$, while our
Eq.~(\ref{eq:Fano_diffTemps_highT1_asymp}) requires $T_{1}\gg T_{2}$.
Second, in Ref.~\citep{Rech2020} the noise becomes negative only
for quasiparticles with $\Delta<1/2$, while Eq.~(\ref{eq:Fano_diffTemps_highT1_asymp})
predicts negative Fano factor for any $\Delta$, cf.~Fig.~\ref{fig:diffTemps_summary}(b,
inset).} Knowing the extent of its negativity enables one to extract the
quasiparticle's scaling dimension $\Delta$, cf.~Eq.~(\ref{eq:Fano_diffTemps_highT1_asymp})
and Fig.~\ref{fig:diffTemps_summary}(b, inset).

The negative excess noise (and thus the negative Fano factor) can be understood from comparing the noise at contact 3 in the presence and in the absence of the tunneling contact. In the absence of tunneling at the QPC, the noise measured at contact 3 is the Johnson--Nyquist noise corresponding to $T_1$. In the presence of tunneling, part of the Johnson--Nyquist noise from the upper edge leaks to the lower edge. Similarly, the noise from the lower edge leaks to the upper edge. The shot noise generated by tunneling can be, to a good approximation, ignored since the temperature imbalance dominates the system. As the upper edge has a higher temperature, overall the noise at contact 3 is reduced. The extent of this reduction is controlled by the intensity of the tunneling processes, i.e., by the scaling dimension $\Delta$. The lower the scaling dimension, the more intense the tunneling is, which correlates with the behavior of $f(\Delta)$ in the inset of Fig.~\ref{fig:diffTemps_summary}(b).

We warn the reader against hastily interpreting the above behaviors in terms of particle statistics or identifying a specific behavior with that of classical particles. As is discussed above and in Appendix~\ref{app:A_edge_theory_background}, the scaling dimension is not always simply related to the quasiparticle statistics. Further, even in the cases when this relation is valid, predictions based on naive models of particles with corresponding statistics can be misleading. For example, the results of Ref.~\cite{Zhang2022} show that the predictions for the noise in the presence of small temperature imbalance are different for the model of free bosons and the model of chiral Luttinger liquid with integer scaling dimensions (that translate into the bosonic statistics of quasiparticles through $\theta=2\pi\Delta$). Therefore, the system's dynamics cannot be described in terms of non-interacting quasiparticles of the respective statistics.

Overall, we argue that the Fano factor is a powerful instrument for
extracting not only the charge but also the scaling dimension of FQH
quasiparticles. In particular, if investigated as a function of the
edge temperatures.

\section{\label{sec:III_Noise_results}General results for thermal noise at
a QPC}

We now proceed to describe our theoretical approach and the
obtained results. Our calculations are valid for \emph{any} FQH edge
theory for which the FQH edges on either side of the QPC can be assigned
a well-defined voltage and temperature. For simplicity, however, we
focus in this section on the Abelian theories. The generalization
to non-Abelian theories is straightforward and is discussed in Appendix~\ref{app:A_edge_theory_background}.

The action of the general Abelian FQH edge is given in terms of $N$
bosonic fields, $\phi_{l},\;i=1,\dots,N$\footnote{The notation we use here is adopted from Refs.~\citep{LBFS_MachZehnder,ShtaSniChe_2014}.
It does not coincide with the standard $K$-matrix-based notation
\citep{Wen2004}, however, is obtained from it via a linear
transformation of the bosonic fields.}
\begin{equation}
S=\frac{1}{4\pi}\int dxdt\sum_{l}\left(-\chi_{l}\partial_{x}\phi_{l}\partial_{t}\phi_{l}-v_{l}\left(\partial_{x}\phi_{l}\right)^{2}\right),\label{eq:Bosonic_Action}
\end{equation}
where $\chi_{l}=\pm1$ and $v_{l}>0$ are the chirality and velocity
of the $l$th mode, respectively. These fields satisfy the commutation
relations
\begin{equation}
\left[\phi_{l}(x,t),\phi_{l'}(x',t')\right]=i\pi\chi_{l}\text{sgn}\left(\xi_{l}-\xi_{l}'\right)\delta_{l,l'},\label{eq:Commutation_relations}
\end{equation}
where $\xi_{l}\equiv x-\chi_{l}v_{l}t$. Density and current operators
along the edge are given by
\begin{equation}
\rho=\frac{1}{2\pi}\sum_{l}q_{l}\partial_{x}\phi_{l};\;j=-\frac{1}{2\pi}\sum_{l}q_{l}\partial_{t}\phi_{l},\label{eq:Explicit_density_current}
\end{equation}
where, as explained in Appendix~\ref{app:A_edge_theory_background},
$q_{l}$ are numbers that define the charge-carrying properties of
the edge modes. They are constrained via the relationship
\[
\sum_{l}\chi_{l}q_{l}^{2}=\nu,
\]
where $\nu$ is the filling factor.

The edge supports quasiparticles of the form
\begin{equation}
\psi_{a}(x,t)\propto e^{i\bm{a}\cdot\bm{\phi}(x,t)},
\end{equation}
where $\bm{\phi}(x,t)\equiv(\phi_{1}(x,t),\dots,\phi_{N}(x,t))$ and
$\bm{a}=(a_{1},\dots,a_{N})$ are vectors composed of the bosonic
fields and real numbers, respectively. The vector $\bm{a}$, while
being real-valued, can only take a discrete set of values reflecting
the set of possible quasiparticles. Such quasiparticles are characterized
by two important quantum numbers: the electric charge

\begin{equation}
Q_{a}=e\sum_{l}\chi_{l}q_{l}a_{l}
\end{equation}
and the scaling dimension
\begin{equation}
\Delta=\frac{1}{2}\sum_{l}a_{l}^{2}.
\end{equation}
The set of allowed quasiparticles must include electrons with $Q=e$.

We model the QPC by a term in the Hamiltonian which describes tunneling
of quasiparticles between the edges,
\begin{align*}
H_{T} & =\sum_{a}\eta_{a}\psi_{a}^{(u)\dagger}\psi_{a}^{(d)}+\text{h.c.}\\
 & \equiv\sum_{a}\left[\eta_{a}A_{a}+\eta_{a}^{*}A_{a}^{\dagger}\right];\;A_{a}\equiv\psi_{a}^{(u)\dagger}\psi_{a}^{(d)},
\end{align*}
with superscripts $u/d$ denoting the upper/lower edge, respectively.
The current operator at the upper/lower drain (labeled 3 and 4 in
Fig.~\ref{fig:QPC_setup}) will be given by\footnote{There is a subtlety present here concerning non-chiral edges. Note
that both the edge current operator $\hat{j}^{(u/d)}$ and the tunneling
current operator involve contributions not only from the downstream
($\chi_{m}=+1$), but also from the upstream ($\chi_{m}=-1$) modes.
The latter, naively, flow away from the respective drain contacts.
We include these contributions as well, motivated by the assumption
of equilibration between different edge modes: eventually any non-equilibrium
current appearing at a mode will end up flowing downstream. We forgo
treating non-chiral edges that do not feature complete equilibration.}
\begin{equation}
\hat{I}_{3/4}=j^{(u/d)}\mp\hat{I}_{T}\label{eq:Edge_Current_Definitions}
\end{equation}
where $j^{(u/d)}$ is the current operator of the unperturbed edge
at the QPC location, cf.~Eq.~(\ref{eq:Explicit_density_current}),
and $I_{T}$ is the tunneling current (i.e., the current that leaves
the upper edge and enters the lower edge). In the operator form, this
is given by
\begin{equation}
\hat{I}_{T}=i\sum_{a}Q_{a}\left[\eta_{a}A_{a}-\eta_{a}^{*}A_{a}^{\dagger}\right].\label{eq:Tunneling_current}
\end{equation}

We wish to calculate the average tunneling current, $I_{T}\equiv\langle\hat{I}_{T}\rangle$,
the auto-correlations at each of the drains, and the cross-correlations
between them. We define the DC noise correlations between drains $i,j$
as
\begin{equation}
S_{i,j}(\omega=0)=\int dt\langle\Delta\hat{I}_{i}(0)\Delta\hat{I}_{j}(t)+\Delta\hat{I}_{j}(t)\Delta\hat{I}_{i}(0)\rangle,\label{eq:Noise_definition}
\end{equation}
where $\Delta\hat{I}_{i}\equiv\hat{I}_{i}-\langle\hat{I}_{i}\rangle$.
The auto-correlations and cross-correlations are not independent,
but satisfy the relation $S_{3,3}+S_{4,4}+2S_{3,4}=2\nu\frac{e^{2}}{h}k_{B}\left(T_{1}+T_{2}\right)$
that links their sum to the Johnson-Nyquist noise, cf.~Appendix~\ref{app:Noise}.
It is, therefore, sufficient to investigate only the auto-correlations
$S_{3,3}$ and $S_{4,4}$.

Without loss of generality, we focus in what follows on the excess
noise measured at drain $3$, $S\equiv S_{3,3}-2\frac{e^{2}}{h}\nu k_{B}T_{1}$.
Defining $\lambda\equiv T_{1}/T_{2}$, we find

\begin{align}
S= & S_{TT}+2S_{0T}\label{eq:Full_Integral_Expressions}\\
S_{TT}= & 4\sum_{a}Q_{a}^{2}G_{a}\mathcal{I}_{1}\left(\frac{Q_{a}V}{\pi k_{B}T_{1}},\frac{1}{\lambda},{2\Delta_{a}}\right),\\
S_{0T}= & -\frac{4i}{\pi}\sum_{a}Q_{a}^{2}G_{a}\mathcal{I}_{2}\left(\frac{Q_{a}V}{\pi k_{B}T_{1}},\frac{1}{\lambda},{2\Delta_{a}}\right)\\
 & -2\sum_{a}Q_{a}^{2}G_{a}\mathcal{I}_{1}\left(\frac{Q_{a}V}{\pi k_{B}T_{1}},\frac{1}{\lambda},{2\Delta_{a}}\right),\nonumber \\
I_{T}= & 2i\sum_{a}Q_{a}G_{a}\mathcal{I}_{3}\left(\frac{Q_{a}V}{\pi k_{B}T_{1}},\frac{1}{\lambda},{2\Delta_{a}}\right),\label{eq:Full_Integral_Expressions_Last}
\end{align}
where we have defined the integrals\footnote{These expressions are valid for $b\leq1$ (equivalently, $\lambda\geq1$).
For general definitions of these integrals, see Appendix~\ref{app:B_Derivations}.}
\begin{align}
\mathcal{I}_{1}\left(a,b,c\right)\equiv\int\limits _{-\infty}^{\infty}dy\frac{\cos{\left(a\left(y-\frac{i\pi}{2}\right)\right)}}{\left[\cosh\left(y\right)i\sinh\left(b\left(y-\frac{i\pi}{2}\right)\right)\right]^{c}},\label{eq:Define_integrals}\\
\mathcal{I}_{2}\left(a,b,c\right)\equiv\int\limits _{-\infty}^{\infty}dy\frac{y\cos{\left(a\left(y-\frac{i\pi}{2}\right)\right)}}{\left[\cosh\left(y\right)i\sinh\left(b\left(y-\frac{i\pi}{2}\right)\right)\right]^{c}},\\
\mathcal{I}_{3}\left(a,b,c\right)\equiv\int\limits _{-\infty}^{\infty}dy\frac{\sin{\left(a\left(y-\frac{i\pi}{2}\right)\right)}}{\left[\cosh\left(y\right)i\sinh\left(b\left(y-\frac{i\pi}{2}\right)\right)\right]^{c}},
\end{align}
and
\begin{equation}
G_{a}\equiv\left|\eta_{a}\right|^{2}\left(\pi k_{B}T_{2}\right)^{4\Delta_{a}-1}\lambda^{2\Delta_{a}-1}\prod_{l}v_{l}^{-2a_{l}^{2}}\label{eq:Effective_Tunneling_Constant}
\end{equation}
is the effective tunneling constant for each quasiparticle. The derivation
of these expressions is given in Appendix~\ref{app:Noise}.

The term $S_{TT}$ arises from self-correlations of the tunneling
current, while $S_{0T}$ represents cross-correlations between the
current $j^{(u/d)}$ flowing along the unperturbed edge and the tunneling
current. The physical mechanism giving rise to these cross-correlations
and its rough relation to the exclusion statistics were described in
Sec.~\ref{sec:II_Main_Results}.

Typically, a single quasiparticle possessing the smallest scaling
dimension $\Delta_{a}=\Delta$ is assumed to dominate the tunneling
processes at the QPC. Denoting its charge $Q_{a}$ as $Q$, we find
the Fano factor
\begin{equation}
F\equiv\frac{S}{2eI_{T}}=-\frac{2Q}{\pi e}\frac{\mathcal{I}_{2}\left(\frac{QV}{\pi k_{B}T_{1}},\frac{T_{2}}{T_{1}},{2\Delta}\right)}{\mathcal{I}_{3}\left(\frac{QV}{\pi k_{B}T_{1}},\frac{T_{2}}{T_{1}},{2\Delta}\right)}.\label{eq:Fano_Integral}
\end{equation}
Note that $F$ does not feature the non-universal tunneling amplitude
$\left|\eta_{a}\right|^{2}$.

\section{\label{sec:IV_experimentally_relevant_scenarios}Experimentally relevant
scenarios}

In this section we explore several parameters' regimes, demonstrating
how the expression for the Fano factor in Eq.~(\ref{eq:Fano_Integral})
enables discrimination between different values of $\Delta$. These
regimes correspond to different cuts of the 2D surface presented in
Fig.~\ref{fig:diffTemps_summary} and illustrate how the three experimental
knobs ($V$, $T_{1}$, and $T_{2}$) affect the Fano factor. We choose
the parameters of these regimes to be compatible with modern experiments
\citep{Dolev2010,Venkatachalam2011,Mills2020,Roosli2021,Pan2020c}.

\subsection{Equal temperatures}

\label{sec:IVA_Equal_temperatures}

The case of equal temperatures was discussed at length in Sec.~\ref{sec:II_Main_Results}.
For $T_{1}=T_{2}\equiv T$, Eq.~\eqref{eq:Fano_Integral} simplifies
to Eq.~\eqref{eq:Fano_sameTemps}, there the Fano factor is equal
to the imaginary part of the digamma function whose argument depends
solely on the parameters $Q$ and $\Delta$, and the ratio $eV/k_{B}T$.
At the high-voltage limit, $eV\gg k_{B}T$, the expression further
simplifies to a linear function of $k_{B}T/eV$, see Eq.~\eqref{eq:Fano_sameTemps_asymp}.

As shown in Fig.~\ref{fig:sameTemps_summary}, this universal function
enables easy extraction of the quantum numbers of interest by plotting
the Fano factor as a function of the ratio $k_{B}T/eV$. The quasiparticle
charge will be given by the intersection of the plot with the $y$
axis, and the scaling dimension by the slope of the curve at low $k_{B}T/eV$.

\subsection{Different temperatures (small V)}

\label{sec:IVB_Small_V}

\begin{figure}
\begin{centering}
\includegraphics[width=1\columnwidth]{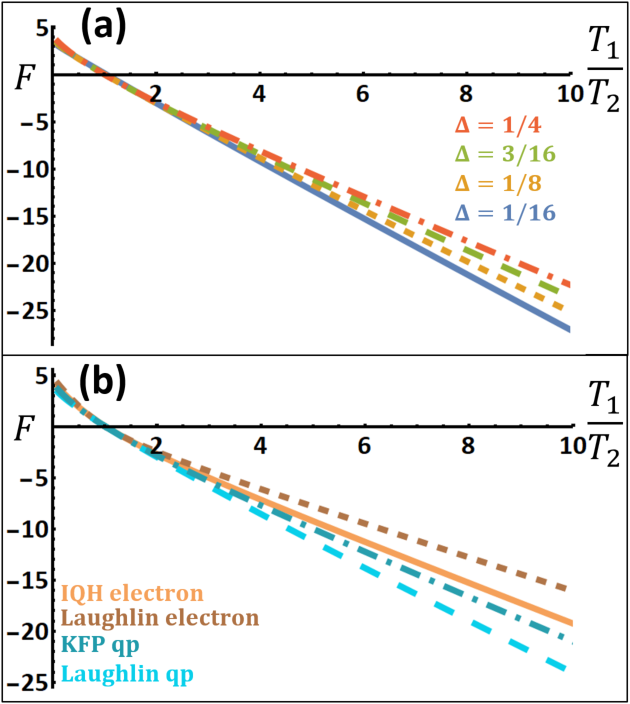}
\par\end{centering}
\caption{\label{fig:diffTemps_Small_V} The Fano factor vs. temperature ratio
$T_{1}/T_{2}$ for small voltage ($eV=0.1\pi k_{B}T_{2}$). Remark
the large $\Delta$-dependent negative values of the Fano factor and
its very weak dependence on the charge $Q$, cf.~Eq.~\eqref{eq:Fano_diffTemps_highT1_asymp}.
(a) Quasiparticles that appear in various candidate theories of $\nu=5/2$
FQH effect: $Q=e/4$ (in all the theories), $\Delta=1/16$ ($K=8$ theory),
$1/8$ (Pfaffian and PH-Pfaffian), $3/16$ (331 state) or $1/4$ (anti-Pfaffian)
\citep{Yang2013}. (b) $\nu=1/3$ Laughlin quasiparticle $(Q=e/3,\Delta=1/6)$,
one of the quasiparticles in the Kane-Fisher-Polchinski (KFP) $\nu=2/3$ fixed-point theory $(Q=e/3,\Delta=1/3)$, non-interacting integer quantum Hall
(IQH) electrons $(Q=e,\Delta=1/2)$, electrons in the $\nu=1/3$ Laughlin
state $(Q=e,\Delta=3/2)$ \citep{WenCLL,KaneFisherPolchinski,KaneFisher}.}
\end{figure}

In this regime, the bias voltage $V$ is kept constant and small
compared with both edge temperatures $T_{1}$ and $T_{2}$. One of
these temperatures is then swept over a substantial range, which in
an experimental setup will be restricted from above by the bulk gap.
The behavior of the Fano factor in this regime is demonstrated in
Fig.~\ref{fig:diffTemps_Small_V}(a) for candidate theories of the
potentially non-Abelian $\nu=5/2$, and in Fig.~\ref{fig:diffTemps_Small_V}(b)
for characteristic quasiparticles at $\nu=1/3$, $2/3$, and $1$.

The Fano factor when $T_{1}=T_{2}$ can be obtained from Eq.~(\ref{eq:Fano_sameTemps});
note that at $V=0$, this will be zero. As $T_{1}$ is increased (decreased),
the Fano factor decreases (increases), becoming strongly negative
(positive). This is consistent with the results of Ref.~\citep{Rech2020},
in which a temperature imbalance leads to noise reduction on the hot
edge, and a noise increase on the cold edge. This regime exhibits
a particularly instructive asymptote where the dominant energy scale
of the system is the temperature of the hot edge, i.e., $\lambda=T_{1}/T_{2}\gg1$.
In this limit, the Fano factor becomes a linear function of the ratio
$k_{B}T_{1}/eV$, cf.~Eq.~\eqref{eq:Fano_diffTemps_highT1_asymp}.
The scaling dimension alone determines the slope via a negative,
monotonously increasing function $f(\Delta)$, cf. the inset of ~Fig.~\ref{fig:diffTemps_summary}(b).
The cold edge temperature $T_{2}$ drops out entirely from the description.
Note the values of $\abs F\gg1$, which appear due to $k_{B}T_{1}\gg eV$.

For physical intuition, we once again appeal to the more familiar
case of Fermi-Dirac distributions. When $k_{B}T_{1}\gg k_{B}T_{2},eV$,
the Fermi sea at the hot edge is so dramatically smeared that any
deformations of the cold edge are comparatively negligible. As such,
the noise will depend solely on $T_{1}$. The tunneling current, meanwhile,
is in the Ohmic limit $I_{T}\propto T_{1}^{4\Delta-2}V$, which leads
to $F\propto V^{-1}$. We note that had we been interested in the
noise measured at contact 4 (cf.~Fig.~\ref{fig:QPC_setup}), belonging
to the colder edge, the respective Fano factor would retain a term
proportional to $T_{2}/T_{1}$.

The monotonicity of the function $f(\Delta)$ makes this regime useful
to differentiate between similar theories with different scaling dimensions.
However, the non-linear form of $f(\Delta)$ means that in this regime
$F$ is highly sensitive for $\Delta<1/2$, less sensitive for $1/2<\Delta<3/2$,
and can hardly discriminate different $\Delta>3/2$.

\subsection{Different temperatures (large voltage)}

\label{sec:IVB_Same_Order}

\begin{figure}
\begin{centering}
\includegraphics[width=1\columnwidth]{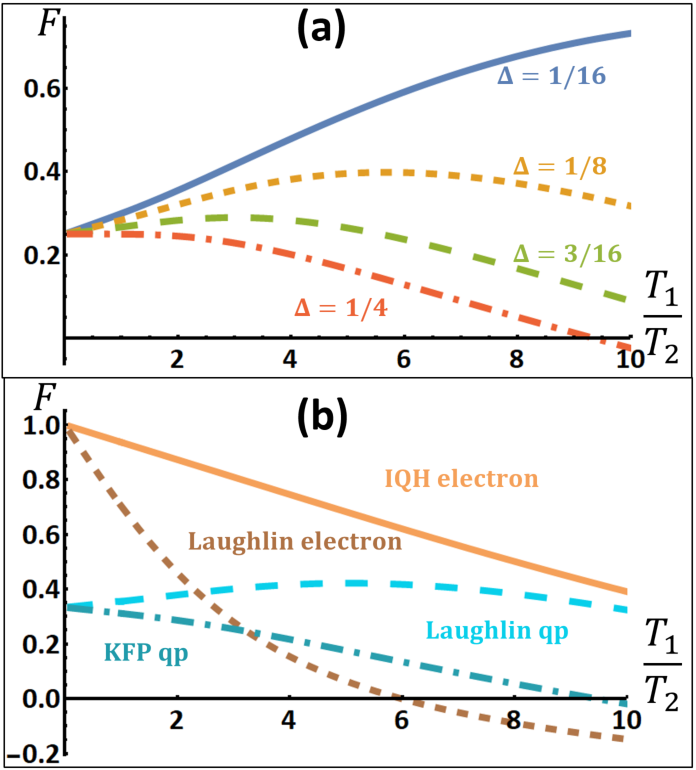}
\par\end{centering}
\caption{\label{fig:diffTemps_SameOrder} The Fano factor vs. temperature ratio
$T_{1}/T_{2}$ for a large voltage ($eV=10\pi k_{B}T_{2}$). Measuring
this dependence, cf.~Eq.~\eqref{eq:Fano_Integral}, one can extract
both the charge $Q$ (which is obtained at $T_{1}/T_{2}\rightarrow0$)
and the scaling dimension $\Delta$ of the tunneling particle (which
dictates the form of the curve). (a) Quasiparticles that appear in
various candidate theories of $\nu=5/2$ FQH effect: $Q=e/4$ (in all the
theories), $\Delta=1/16$ ($K=8$ theory), $1/8$ (Pfaffian and PH-Pfaffian),
$3/16$ (331 state) or $1/4$ (anti-Pfaffian) \citep{Yang2013}. (b)
$\nu=1/3$ Laughlin quasiparticles $(Q=e/3,\Delta=1/6)$, one of the
quasiparticles in the Kane-Fisher-Polchinski (KFP) $\nu=2/3$ fixed-point theory
$(Q=e/3,\Delta=1/3)$, non-interacting integer quantum Hall (IQH)
electrons $(Q=e,\Delta=1/2)$, electrons in the $\nu=1/3$ Laughlin
state $(Q=e,\Delta=3/2)$ \citep{WenCLL,KaneFisherPolchinski,KaneFisher}.}
\end{figure}

Another useful regime exists when $eV\gg k_{B}T_{2}$, while $k_{B}T_{1}$
takes any value between them. We present the data for candidate theories
at $\nu=5/2$ and for characteristic quasiparticles at $\nu=1/3$,
$2/3$, and $1$ in Fig.~\ref{fig:diffTemps_SameOrder}, panels (a)
and (b) respectively.

As long as $k_{B}T_{1}\ll eV$, the Fano factor obeys Eq.~(\ref{eq:Fano_diffTemps_highV_asymp}),
in particular giving the charge $Q$ at the limit $T_{1}\rightarrow0$.
The dependence on $\Delta$ in this regime is linear, which guarantees
the same sensitivity over the whole range of $\Delta$.

For $k_{B}T_{1}\lesssim eV$, the analytical understanding of the
behavior is lacking. However, Fig.~\ref{fig:diffTemps_SameOrder}
shows that this regime has its own distinctive features. Note that
$F$ can become negative. We find that it always becomes negative
at $k_{B}T_{1}\simeq eV$, with the exact location of the crossover
point determined by the scaling dimension. This agrees with the intuition
of the $k_{B}T_{1}\gg eV$ regime, cf.~Sec.~\ref{sec:IVB_Small_V}.

\section{\label{sec:V_Discussion}Discussion}

In the above sections, we have described a method for determining the scaling dimension of quantum Hall quasiparticles and analyzed some experimentally relevant regimes. At the same time, it is important to understand what information is encoded in the scaling dimension. It is also important to be aware of the experimental subtleties that may affect the applicability of the above considerations. We discuss these issues below.

In the fully chiral edges (both Abelian and non-Abelian), the scaling
dimension is universal as it is related to the quasiparticle statistics, cf.~Appendix~\ref{app:A_edge_theory_background}. Non-chiral edges do not feature this robustness: interactions
between counterpropagating edge modes can lead to a change in the
scaling dimension \citep{KaneFisherPolchinski,KaneFisher,Wang2013a,Sun2020d}.
Further, edge reconstruction can add counterpropagating modes to chiral
edges \citep{Rosenow2002,Wang2013a} and thus facilitate a change
in the scaling dimension. Yet even under these circumstances the scaling
dimension remains an important quantity to measure. First, the scaling
dimension reflects the properties of the edge \emph{including} the
reconstructions and intermode interactions. Second, the scaling dimension
of the quasiparticle that dominates tunneling is typically larger
in the reconstructed theory. Therefore, measuring a specific scaling
dimension excludes underlying non-reconstructed theories, where the
scaling dimension is bigger than the one measured.

Another mechanism undermining the universality of the scaling dimension
is the electrostatic (screened Coulomb) interactions in the vicinity
of the QPC~\citep{Papa2004,Yang2013}. These may renormalize the
scaling dimension in the vicinity of the QPC, so that it does not reflect
the properties of the quasiparticles away from the QPC. One can minimize
these interaction effects by designing the device such that counter
propagating modes are close to each other only at short length $l_{{\rm int}}$
near the QPC. At low energies such that $v/E\gg l_{{\rm int}}$ ($v$ is the characteristic edge velocity) the
Coulomb interaction will not affect the scaling dimension. The results of Ref.~\cite{Bartolomei2020} suggest that it is indeed possible to have experiments, where the electrostatic interactions in the vicinity of the QPC do not play a major role.

The above non-universalities may affect the interpretation of the extracted scaling dimension. However, they do not affect the validity of our method. Below, we discuss subtleties that may be present in realistic experimental setups and may require modifications to the proposed method.

The considerations of this paper assume that each
edge is at equilibrium. However, edges featuring counterpropagating
modes may not be at equilibrium \citep{Nosiglia2018a,Aharon-Steinberg2019,Park2019,Spanslatt2019,Ma2019,Simon2020,Asasi2020}.
Some types of non-equilibrium may be tolerated. For example, the $\nu=2/3$
edge can have temperature gradients along the edge while \emph{locally}
all the modes have the same temperature \citep{Nosiglia2018a,Park2019,Spanslatt2019}.
This type of non-equilibrium can be taken into account simply: the
physics at the QPC is described by the local temperature. As long
as the temperature at the QPC can be estimated independently, the
scaling dimension can be extracted.

On the other hand, if counterpropagating
modes interact very weakly, they can be out of equilibrium even locally,
invalidating the assumptions of the present work. Such non-equilibrium
will, however, lead to a non-quantized Hall conductance for charged
counterpropagating modes \citep{KaneFisherPolchinski,KaneFisher}.
The equilibration properties of counterpropagating neutral modes can
be investigated by measuring the excess noise of a single edge (with
no tunneling at a QPC) \citep{Nosiglia2018a,Park2019,Spanslatt2019,Park2020b,Melcer2021,Kumar2021b, [See also ] Spanslatt2020}.

Finally, one could expect complications due to interfaces between the Ohmic contacts and the quantum Hall edges. Indeed, the transport properties (e.g., conductance) of the Luttinger liquid depend crucially on the nature of its interface with external leads (see, e.g., Refs.~\citep{Alekseev1996,Kloss2018}). Quantum Hall edges may also experience such non-universal effects \citep{Oreg1995,Slobodeniuk2013,Spanslatt2021}, implying the need to employ other methods for probing the noise generated at a QPC, e.g., that of Ref.~\citep{Tikhonov2016}. However, we expect this to be unnecessary. Such non-universalities should have no influence on the observations as long as the charge transport along the edge channels is fully chiral (i.e., for the edges where all modes have the same chirality, as well as for generic edges with modes of different chiralities in the regime of strong equilibration~\citep{Nosiglia2018a}). All the current and noise generated at the QPC must then reach the respective Ohmic contact, as they cannot be reflected back at the interface.

\section{\label{sec:VI_Conclusion}Conclusions}

In this paper we have proposed a method for determining the scaling
dimension of quantum Hall quasiparticles based on the measurements
of the noise generated at a QPC. The method relies on analyzing the
dependence of the Fano factor on the bias voltage and the temperatures
of the quantum Hall edges. We expect the extraction of the scaling
dimension via the proposed method to be as robust as the extraction
of the quasiparticle charge from the Fano factor.

While our method is expected to enable reliable extraction of the
scaling dimension and excludes some non-universal effects, it is important
to realize that the scaling dimension itself may not be universal. Yet even when the scaling
dimension is non-universal, it remains an important quantity that characterizes the dynamics of the system. For fully chiral edges (both Abelian and non-Abelian), the scaling
dimension is universal, as it is related to the quasiparticle statistics.

\begin{acknowledgements}

\section*{Acknowledgements}

We thank Amir Rosenblatt, Sofia Konyzheva, Tomer Alkalay, and Moty
Heiblum for useful discussions and Christian Spånslätt, Gu Zhang, Alexander Mirlin, and Igor Gornyi for insightful comments on the manuscript. K.S. acknowledges funding by the Deutsche
Forschungsgemeinschaft (DFG, German Research Foundation) -- Projektnummer
277101999 -- TRR 183 (project C01), Projektnummer GO 1405/6-1, Projektnummer
MI 658/10-2, and by the German-Israeli Foundation Grant No. I-1505-303.10/2019.
The work at Weizmann was partially supported by grants from the ERC
under the European Union’s Horizon 2020 research and innovation programme
(grant agreements LEGOTOP No. 788715 and HQMAT No. 817799), the DFG
(CRC/Transregio 183, EI 519/7-1), the BSF and NSF (2018643), the ISF
Quantum Science and Technology (2074/19). NS was supported by the Clore Scholars Programme.

\end{acknowledgements} \bibliography{bibliography}

\appendix

\newpage{}

\section{\label{app:A_edge_theory_background}Background on quantum Hall edge
theory}

In this section we remind the reader some well-known theoretical results
which apply to all existing quantum Hall edge models (integer and
fractional, Abelian and non-Abelian, fully chiral or featuring counterpropagating
modes). We only focus here on the aspects relevant for the present
work, in particular, the role of the scaling dimension in describing
the edge properties. For a more comprehensive summary of the theory
in the notation that is close to the notation used in the present
paper, see Ref.~\citep[Sec. III]{LBFS_MachZehnder} or Ref.~\citep[Sec. IV]{ShtaSniChe_2014}.
Among other details, these references explain the relation between
the standard $K$-matrix-based notation for Abelian theories~\citep{WenReview}
and the present notation.

\subsection{\label{app:A.5.a_general_edge_model}The structure of a general edge
model}

\begin{figure*}
\begin{centering}
\includegraphics[width=1\textwidth]{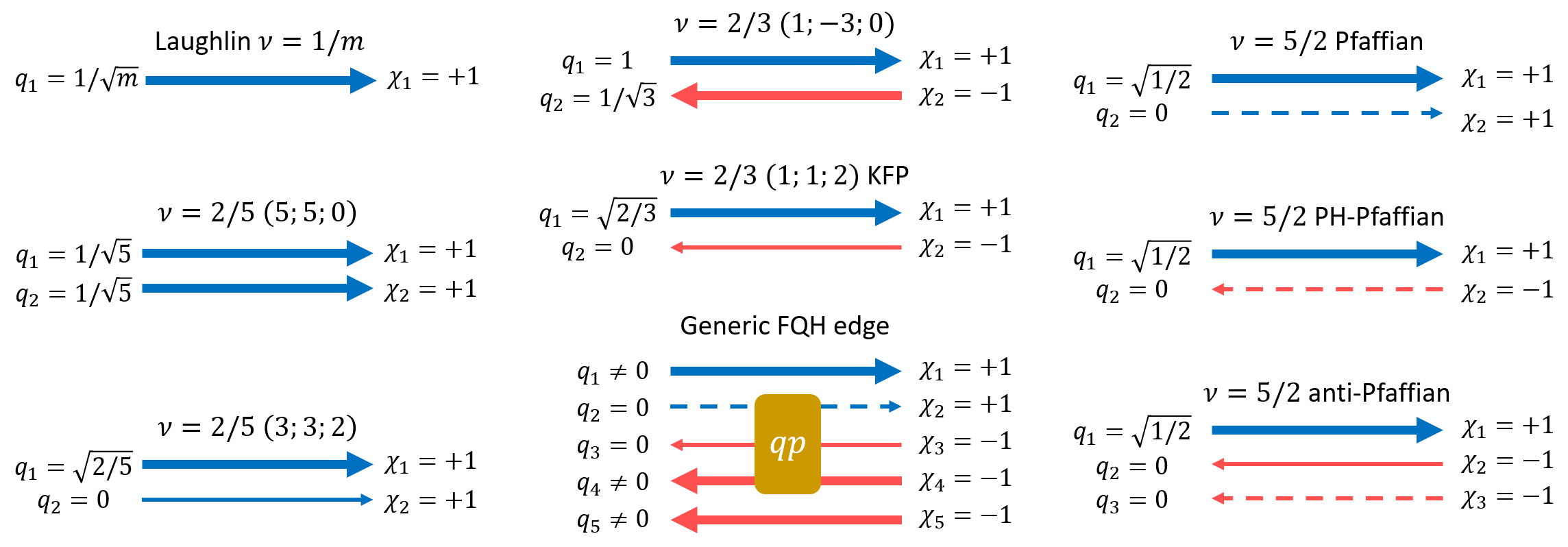}
\par\end{centering}
\caption{\label{fig:QH_edge_structures}Examples of various quantum Hall edge
structures. Note that the same filling factor $\nu$ may admit multiple
edge structures. Each edge structure is characterized by a number
of gapless modes. Each mode is chiral (rightmoving, $\chi=+1$ {[}blue
arrows{]}, or leftmoving, $\chi=-1$ {[}red arrows{]}). Modes can
be charged (thick lines) and neutral (thin). Further, they can be
Abelian (solid) or support non-Abelian excitations (dashed). Quasiparticles
(\emph{qp}) that can tunnel between edges of one FQH sample generically
combine degrees of freedom from several modes.}
\end{figure*}

The behavior of quantum Hall edges is theoretically described by the
chiral Luttinger liquid theory and its non-Abelian generalizations
\citep{WenCLL,FrohlichAnomaly,WenReview,Bieri2011}. A quantum Hall
edge may consist of an arbitrary number $N$ of edge modes. Each mode
has a direction of propagation (chirality) that we denote $\chi_{l}=\pm1$,
$l=1,...,N$, cf.~Fig.~\ref{fig:QH_edge_structures}. Some modes
contribute to the electric transport by carrying charged excitations,
while others may carry only neutral excitations. The size of a mode's
contribution to the electric transport can be encoded in numbers $q_{l}\geq0$.
When a mode does not carry charged excitations, $q_{l}=0$. If the
edge mode is at equilibrium and its electrostatic potential is $V$,
it carries current $I_{l}=\frac{e^{2}}{h}q_{l}^{2}\chi_{l}V$. The
sign of $I_{l}$ reflects the current direction, i.e., the mode's
chirality. Fixing the Hall conductance to $\nu e^{2}/h$, therefore,
requires $\sum_{l}q_{l}^{2}\chi_{l}=\nu$. Each mode has a velocity
$v_{l}>0$ with which its excitations propagate.\footnote{Experts may note that electrostatic interactions between different
modes lead to terms in the Hamiltonian that require defining the velocity
matrix instead of assigning velocities to each of the modes (the latter
corresponds to the velocity matrix being diagonal). For example, this
is the case for electrostatic interactions between the microscopic
$1$ and $1/3$ edge modes in $(1;-3;0)$ theory for $\nu=2/3$, cf.~Refs.~\citep{KaneFisherPolchinski,KaneFisher}.
We point out that the positive-definite velocity matrix can always
be diagonalized while also preserving the ``chirality matrix'' diagonal,
aligning the formalism with our notation. The effect of the interactions
is then encoded in the contributions $\tilde{q}_{l}$ of the new modes
to the electric transport. In other words, the new non-interacting
modes will contribute differently to the electric transport than the
original interacting microscopic modes.}

The bulk and the edge of a quantum Hall sample can host local quasiparticles.
At the edge, for each quasiparticle one can define the second-quantized
quasiparticle operator $\psi_{\mathrm{qp}}(x,t)$, where $x$ is the
position along the edge and $t$ is the time. The quasiparticle might
be associated with one mode only. However, in general, a quasiparticle
is distributed over several modes: $\psi_{\mathrm{qp}}(x,t)=\prod_{l}O_{l}(x,t)$,
where $O_{l}$ is an operator belonging to mode $l$. The distribution
can be quantified by a set of numbers $h_{l}\geq0$, reflecting the
scaling properties of operators $O_{l}$. For the modes contributing
to the electric transport, one defines $a_{l}=\sqrt{h_{l}}$. The
electric charge of the quasiparticle can then be expressed as
\begin{equation}
Q=e\bm{q}\circ\bm{a}=e\sum_{l}q_{l}\chi_{l}a_{l},\label{eq:qp_charge_definition}
\end{equation}
where $e$ is the electron charge. It is also convenient to define
the scaling dimension
\begin{equation}
\Delta=\frac{1}{2}\sum_{l}h_{l}\label{eq:scaling_dimension_definition}
\end{equation}
and the conformal spin\footnote{We note that while the conformal spin $s$ of a quasiparticle can
be expressed in terms of the $K$-matrix, the scaling dimension $\Delta$
does not have such an expression in general, cf.~Ref.~\citep[Sec. III]{LBFS_MachZehnder}.}
\begin{equation}
s=\frac{1}{2}\sum_{l}\chi_{l}h_{l}.\label{eq:conformal_spin_definition}
\end{equation}

When the edge is at equilibrium, characterized by electrostatic potential
$V$ and temperature $T$, one can write the self-correlation function
of a quasiparticle on the edge as
\begin{multline}
\langle\mathcal{T}\psi_{\mathrm{qp}}^{\dagger}(x,t)\psi_{\mathrm{qp}}(0,0)\rangle\\
=\prod_{l}\left[\left(\frac{\pi k_{B}T}{i\hbar v_{l}\sinh\pi k_{B}Tt_{l}/\hbar}\right)^{h_{l}}\exp\left(\frac{i}{\hbar}eq_{l}\chi_{l}a_{l}Vt_{l}\right)\right],\label{eq:qpdag-qp_correlation_function}
\end{multline}
and
\begin{multline}
\langle\mathcal{T}\psi_{\mathrm{qp}}(x,t)\psi_{\mathrm{qp}}^{\dagger}(0,0)\rangle\\
=\prod_{l}\left[\left(\frac{\pi k_{B}T}{i\hbar v_{l}\sinh\pi k_{B}Tt_{l}/\hbar}\right)^{h_{l}}\exp\left(-\frac{i}{\hbar}eq_{l}\chi_{l}a_{l}Vt_{l}\right)\right],\label{eq:qp-qpdag_correlation_function}
\end{multline}
where $t_{l}=t-\chi_{l}x/v_{l}-i0^{+}\sgn\,t$ with $0^{+}$ being
the infinitesimally small positive number and $\mathcal{T}$ stands
for the time ordering of the operators.

Note how the above correlation functions reflect the quasiparticle
distribution over several edge modes. The correlation function in
Eq.~(\ref{eq:qp-qpdag_correlation_function}) represents the quantum
amplitude of putting a quasiparticle on the edge at~$t=0$ and extracting
it from a different location at a later time~$t>0$. The size of this
amplitude decays exponentially for $T\neq0$ (power-law when $T\rightarrow0$)
with the distance from the expected quasiparticle location on each
edge mode: $x_{\mathrm{expected}}^{(j)}=\chi_{l}v_{l}t$. However,
the expected location $x_{\mathrm{expected}}^{(l)}$ is different
for different modes~$l$. This reflects that once the quasiparticle
is put onto the FQH edge, different parts of it propagate independently.
This is reminiscent of spin-charge separation in Luttinger liquids:
once an electron is injected into a Luttinger liquid, its spin and
charge propagate with different velocities \citep{Schulz1995}.

This concludes the minimalistic introduction into the structure of
a general quantum Hall edge model. Next, in Sec.~\ref{app:A.5.b_qp_properties_analysis},
we focus on discussing the main properties of the quasiparticles,
namely, the charge $Q$, the conformal spin $s$, the scaling
dimension $\Delta$, and their connection to the quasiparticle statistics
and to the setup of Fig.~\ref{fig:QPC_setup}.

\subsection{\label{app:A.5.b_qp_properties_analysis}Properties of the FQH quasiparticles
and their connection to the setup in Fig.~\ref{fig:QPC_setup}}

In order to explain the meaning of the quasiparticle scaling dimension
$\Delta$ in Eq.~(\ref{eq:scaling_dimension_definition}) and conformal
spin $s$ in Eq.~(\ref{eq:conformal_spin_definition}), as well as
their connection to the statistics, we will analyze the quasiparticle
correlation functions in Eqs.~(\ref{eq:qpdag-qp_correlation_function}--\ref{eq:qp-qpdag_correlation_function}).
Consider first the same-position correlation function:

\begin{equation}
\langle\psi_{\mathrm{qp}}(0,t>0)\psi_{\mathrm{qp}}^{\dagger}(0,0)\rangle\propto\frac{e^{-iQVt/\hbar}}{\left(\sinh\pi k_{B}Tt/\hbar\right)^{2\Delta}}.\label{eq:same-position_corr_func}
\end{equation}
The oscillating exponential confirms the meaning of the quasiparticle
charge $Q$ defined in Eq.~(\ref{eq:qp_charge_definition}): creating
a quasiparticle adds energy $QV$ to the edge. The scaling dimension
$\Delta$ controls the decay of self-correlations with time.

Consider now the correlation of quasiparticles at the same time but
different positions. These can be obtained from Eqs.~(\ref{eq:qpdag-qp_correlation_function}--\ref{eq:qp-qpdag_correlation_function})
by taking the limit $t\rightarrow0$ while keeping $t>0$ and using
the translational invariance of the correlation functions:
\begin{multline}
\langle\psi_{\mathrm{qp}}^{\dagger}(x,0)\psi_{\mathrm{qp}}(0,0)\rangle\\
\propto\prod_{l}\left[-i\sinh\frac{\pi k_{B}T\left(i0^{+}+\chi_{l}x/v_{l}\right)}{\hbar}\right]^{-h_{l}},
\end{multline}

\begin{multline}
\langle\psi_{\mathrm{qp}}(0,0)\psi_{\mathrm{qp}}^{\dagger}(x,0)\rangle\\
\propto\prod_{l}\left[-i\sinh\frac{\pi k_{B}T\left(i0^{+}-\chi_{l}x/v_{l}\right)}{\hbar}\right]^{-h_{l}}.
\end{multline}
Ignoring the infinitesimal imaginary part, one sees that the two expressions
are connected by a factor $\prod_{l}(-1)^{-h_{l}}$. The infinitesimal
imaginary part prescribes what root of unity one should take in each
factor depending on the chirality $\chi_{l}$, so that
\begin{equation}
\langle\psi_{\mathrm{qp}}(0,0)\psi_{\mathrm{qp}}^{\dagger}(x,0)\rangle=e^{-2i\pi s\,\sgn\,x}\langle\psi_{\mathrm{qp}}^{\dagger}(x,0)\psi_{\mathrm{qp}}(0,0)\rangle.
\end{equation}
This suggests that exchanging two quasiparticles produces the statistical
factor $e^{i\theta}$ with $\theta=\pm2\pi s$. Therefore, the conformal
spin in Eq.~(\ref{eq:conformal_spin_definition}) reflects the statistics
of the quasiparticles. The last statement is accurate only for Abelian
quasiparticles. If the quasiparticles possess non-Abelian statistics,
this requires at least four quasiparticles to be manifest and cannot
be seen via two-point correlation functions. Therefore, the conformal
spin $s$ only captures the Abelian part of the statistics.

When the edge contains modes of one chirality only (all $\chi_{l}=\chi$),
then $s=\chi\Delta$, as can be seen from Eqs.~(\ref{eq:scaling_dimension_definition}--\ref{eq:conformal_spin_definition}).
In this sense, measuring the scaling dimension allows one to infer
the quasiparticle statistics. We emphasize once more that this statement
is only exactly valid for edges that are fully chiral and feature no non-Abelian
quasiparticles. However, even when the correspondence between the
statistics and the scaling dimension does not hold, the scaling dimensions
of different quasiparticles in the theory are an important property
of the edge theory and are valuable to measure.

Tunneling experiments as in Fig.~\ref{fig:QPC_setup} enable access
to the scaling dimension. Since the tunneling happens only at one
point on each edge, such experiments can be described in terms of
the same-position correlation functions such as in Eq.~(\ref{eq:same-position_corr_func}).
In fact, when the tunneling is weak, only two-point correlation functions,
containing one creation and one annihilation operator for a quasiparticle,
appear in the calculation, cf.~Appendix~\ref{app:B_Derivations}.
Therefore, nothing but the quasiparticle charge and scaling dimension
can be extracted from such experiments (at least, at weak tunneling).\footnote{This statement, applies in particular to the experiment of Ref.~\citep{Bartolomei2020}.
The experiment measures a quantity related to the scaling dimension
of the Laughlin quasiparticle, as is explicitly explained in the theory
proposal, Ref.~\citep{Rosenow2016}, the experiment is based on.
The relation to the statistics holds because of the simple structure
of the Laughlin state and the experimentalists being able to avoid
non-universal behavior of the device (such as edge reconstruction
or strong interactions of different modes across the QPC).}

It is important to point out that a quantum Hall edge hosts many types
of quasiparticles, all of which can contribute to tunneling.\footnote{In the simplest case of the Laughlin $\nu=1/3$ state, these would
be the Laughlin quasiparticles, but also the agglomerates of two,
three and more quasiparticles bunched together, cf.~Ref.~\citep{Chung2003}.} It can be argued (and confirmed numerically \citep{HuntingtonCheianov_TunnAmplitudeMonteCarlo})
that at sufficiently low energies the quasiparticle with the smallest
scaling dimension (whose correlations decay the slowest in time, cf.~Eq.~(\ref{eq:same-position_corr_func}))
dominates the tunneling processes. In the setup of Fig.~\ref{fig:QPC_setup}(a),
therefore, one expects to measure the charge and the scaling dimension
of the fractional quasiparticle whose scaling dimension is the smallest.
This quasiparticle is also called the most relevant quasiparticle.
There are theories in which several quasiparticles possess the smallest
scaling dimension, cf.~Refs.~\citep{KaneFisherPolchinski,ShtaSniChe_2014,HalperinRosenow_2007_AntiPfaffian,Nayak_2007_AntiPfaffian}.
In this case the contribution of all such quasiparticles has to be
taken into account.

In the setup of Fig.~\ref{fig:QPC_setup}(b), fractional quasiparticles
cannot tunnel; only electrons and agglomerates of electrons can. Therefore,
the setup of Fig.~\ref{fig:QPC_setup}(b) enables measuring the scaling
dimension of the electron. We stress that this is a non-trivial measurement,
also characterizing the FQH edge and the interacting nature of the
state. On an \emph{integer} quantum Hall edge, the electron is expected
to have the scaling dimension of $\Delta=1/2$. Since the IQH edge
is fully chiral, this is related to the fermionic statistics of the
electron: $\theta=2\pi s=2\pi\Delta=\pi$, $e^{i\theta}=-1$. By contrast,
in the Laughlin $\nu=1/3$ state, the electron is expected to have
$\Delta=3/2$. Laughlin edge is also chiral, so the statistics is still
manifestly fermionic: $\theta=2\pi s=2\pi\Delta=3\pi$, $e^{i\theta}=-1$.
However, the dynamical behavior of electrons is altered due to the
strongly-correlated nature of the state, which results in a different
scaling dimension.

\newpage{}

\begin{widetext}
\section{\label{app:B_Derivations}Theory derivations}
\global\long\def\theequation{B\arabic{equation}}%
\setcounter{equation}{0}

\subsection{\label{app:CorrelationFunctions}Required correlation functions}

Now we want to calculate various two point functions. From the action in Eq.~\eqref{eq:Bosonic_Action} we obtain
\begin{equation}
\left\langle \phi_{l}(x,t)\phi_{l}(x',t')\right\rangle =\log\left[\frac{\sinh\left(-\frac{\chi_{l}\pi k_{B}T}{v_{l}}\left(\xi_{l}-\xi{}_{l}^{\prime}+i\chi_{l}\delta\right)\right)}{\sinh\left(\frac{\pi k_{B}T}{iv_{l}}\delta\right)}\right],\label{eq:Two_Point_Function}
\end{equation}
where $\xi_{l}\equiv x-\chi_{l}v_{l}t$, $\delta$ is a short distance
cutoff, and we are working in units where $\hbar=1$. This leads to

\begin{equation}
\left\langle \psi_{a}^{\dagger(u/d)}(x,t)\psi_{a'}^{(u/d)}(x',t')\right\rangle =\delta_{a,a'}\prod_{l}\left[\frac{\left(\frac{\pi k_{B}T_{1/2}}{iv_{l}^{(u/d)}}\right)}{\sinh\left(-\frac{\chi_{l}\pi k_{B}T_{1/2}}{v_{l}^{(u/d)}}\left(\xi_{l}-\xi{}_{l}^{\prime}-i\chi_{l}\delta\right)\right)}\right]^{a_{m}^{2}}e^{-ieq_{l}a_{l}V_{1/2}\frac{(\xi_{l}-\xi{}_{l}^{\prime})}{v_{l}^{(u/d)}}},\label{eq:Two_Vertex_Function}
\end{equation}
where $u/d$ denotes a quasiparticle operator for the upper/lower
edge, and $V_{1/2}$ and $T_{1/2}$ are the electrostatic potentials
and the temperatures of the edges, respectively. Subsequently,
\begin{equation}
\left\langle A_{a}^{\dagger}(t)A_{a'}(0)\right\rangle =\delta_{a,a'}\prod_{l}\left[\frac{\left(\frac{\pi k_{B}T_{1}}{iv_{l}^{(u)}}\frac{\pi k_{B}T_{2}}{iv_{l}^{(d)}}\right)}{\sinh\left(\pi k_{B}T_{1}\left(t-i\varepsilon\right)\right)\sinh\left(\pi k_{B}T_{2}\left(t-i\varepsilon\right)\right)}\right]^{a_{m}^{2}}e^{ie\chi_{l}q_{l}a_{l}Vt},\label{eq:Two_Tunnel_Operator_Function}
\end{equation}
where $V\equiv V_{1}-V_{2}$ is the bias drop between the edges, and
$\varepsilon$ is a short time cutoff. Furthermore, using the definition
of $j^{(u)}(0)$ in Eq.~\eqref{eq:Explicit_density_current}, we
obtain

\begin{equation}
\left\langle j^{(u/d)}(0)j^{(u/d)}(t)\right\rangle =\left(\frac{1}{2\pi}\right)^{2}\nu\frac{\left(\pi k_{B}T_{1/2}\right)^{2}}{\sinh^{2}\left(\pi k_{B}T_{1/2}t\right)}.\label{eq:Thermal_Current_Correlation}
\end{equation}

The final correlation function we need is

\begin{multline}
\left\langle \Delta j^{(u/d)}(x,t)\psi_{a}^{\dagger}(x',t')\psi_{a'}(x'',t'')\right\rangle =\delta_{a,a'}\left\langle \psi_{a}^{\dagger}(x',t')\psi_{a}(x'',t'')\right\rangle \frac{Q\pi k_{B}T_{1/2}\chi_{l}}{2\pi i}\times\\
\left[\coth\left(-\frac{\chi_{l}\pi k_{B}T_{1/2}}{v_{l}}(\xi_{l}-\xi{}_{l}^{\prime}-i\chi_{l}\delta)\right)-\coth\left(-\frac{\chi_{l}\pi k_{B}T_{1/2}}{v_{l}}\left(\xi_{l}-\xi_{l}^{\prime\prime}-i\chi_{l}\delta\right)\right)\right],\label{eq:Current_and_vertex}
\end{multline}
which recreates Eq. (A4) in Ref.~\citep{ShtaSniChe_2014}.

\subsection{\label{app:Noise}Noise}

We define the DC noise correlations between drains $i,j$ as
\begin{equation}
S_{i,j}(\omega=0)=\int dt\langle\Delta\hat{I}_{i}(0)\Delta\hat{I}_{j}(t)+\Delta\hat{I}_{j}(t)\Delta\hat{I}_{i}(0)\rangle,
\end{equation}
where $\Delta\hat{I}_{i}\equiv\hat{I}_{i}-\langle\hat{I}_{i}\rangle$
(cf. Eq.~\eqref{eq:Noise_definition} of the main text). From Eqs.~\eqref{eq:Edge_Current_Definitions}
and \eqref{eq:Tunneling_current}, we obtain to the leading order
in the tunneling amplitude $\eta$
\begin{align*}
\left\langle \hat{I}^{(3/4)}\right\rangle = & \left\langle j^{(u/d)}\mp\hat{I}_{T}\right\rangle =\left\langle j^{(u/d)}\right\rangle \\
\Delta\hat{I}^{(3/4)}= & \Delta j^{(u/d)}\mp\hat{I}_{T},
\end{align*}
where $\Delta j^{(u/d)}=j^{(u/d)}-\left\langle j^{(u/d)}\right\rangle $.
Auto-correlations and cross-correlations will hence be given by
\begin{align}
S_{33/44} & =S_{00}^{(u/d)}+S_{0T}^{(u/d)}+S_{T0}^{(u/d)}+S_{TT},\label{eq:AllFourCorrelations}\\
S_{34/43} & =-S_{TT}-S_{0T}^{(u/d)}-S_{T0}^{(d/u)},
\end{align}
where we define
\begin{align}
S_{TT}= & \int dt\left\langle \{\hat{I}_{T}(0),\hat{I}_{T}(t)\}\right\rangle ,\nonumber \\
S_{0T}^{(u/d)}=S_{T0}^{(u/d)}= & \mp\int dt\left\langle \{\Delta j^{(u/d)}(0),\hat{I}_{T}(t)\}\right\rangle ,\label{eq:Define_noise_terms}\\
S_{00}^{(u/d)}= & \int dt\left\langle \{\Delta j^{(u/d)}(0),\Delta j^{(u/d)}(t)\}\right\rangle ,\nonumber
\end{align}
and we used
\begin{equation}
\left\langle j^{(u)}(0)j^{(d)}(t)\right\rangle =0.
\end{equation}
So the entirety of the noise correlations are described by five different
terms, which boil down to three independent calculations: $S_{TT}$,
$S_{0T}^{(u/d)}$, $S_{00}^{(u/d)}$.

\subsubsection{$S_{00}$}

This term is derived directly from Eq.~\eqref{eq:Thermal_Current_Correlation},
\begin{multline}
S_{00}^{u/d}=\int dt\left\langle \{\Delta j^{(u/d)}(0),\Delta j^{(u/d)}(t)\}\right\rangle \\
=2\left(\frac{1}{2\pi}\right)^{2}\nu\pi k_{B}T_{1/2}\int dt\frac{\pi k_{B}T_{1/2}}{\sinh^{2}\left(\pi k_{B}T_{1/2}t\right)}=\left.\left(\frac{1}{2\pi}\right)\nu k_{B}T_{1/2}\coth\left(x\right)\right|_{-\infty}^{\infty}=\frac{\nu}{\pi}k_{B}T_{1/2}.
\end{multline}
In this manner we see that for each edge, (up to restoration of $e,\hbar$),
$S_{00}$ just gives the Johnson-Nyquist noise. As we are interested
in excess noise, i.e., noise measured beyond the Johnson-Nyquist noise,
this is subtracted from the total noise contribution we seek in the
main text.

\subsubsection{$S_{0T}=S_{T0}$}

We define $S_{0T}\equiv S_{0T}^{(u)}$. Plugging Eq.~\eqref{eq:Tunneling_current}
into Eq.~\eqref{eq:Define_noise_terms}, we use the Kubo formula
to obtain
\begin{align*}
S_{0T} & =-\int\limits _{-\infty}^{\infty}dt\left\langle j^{(u)}(0)\hat{I}_{T}(t)+\hat{I}_{T}(t)j^{(u)}(0)\right\rangle \\
 & \rightarrow i\int\limits _{-\infty}^{\infty}dt\int\limits _{-\infty}^{t}d\tau\left\langle j^{(u)}(0)\left[\hat{I}_{T}(t),\sum_{a}\left(\eta_{a}A_{a}(\tau)+\eta_{a}^{*}A_{a}^{\dagger}(\tau)\right)\right]+\left[\hat{I}_{T}(t),\sum_{a}\left(\eta_{a}A_{a}(\tau)+\eta_{a}^{*}A_{a}^{\dagger}(\tau)\right)\right]j^{(u)}(0)\right\rangle \\
 & =-\sum_{a}\left|\eta_{a}\right|^{2}Q_{a}\int\limits _{-\infty}^{\infty}dt\int\limits _{-\infty}^{t}d\tau\left\langle j^{(u)}(0)\left(A_{a}(t)A_{a}^{\dagger}(\tau)-A_{a}^{\dagger}(t)A_{a}(\tau)-A_{a}^{\dagger}(\tau)A_{a}(t)+A_{a}(\tau)A_{a}^{\dagger}(t)\right)\right\rangle \\
 & -\sum_{a}\left|\eta_{a}\right|^{2}Q_{a}\int\limits _{-\infty}^{\infty}dt\int\limits _{-\infty}^{t}d\tau\left\langle \left(A_{a}(t)A_{a}^{\dagger}(\tau)-A_{a}^{\dagger}(t)A_{a}(\tau)-A_{a}^{\dagger}(\tau)A_{a}(t)+A_{a}(\tau)A_{a}^{\dagger}(t)\right)j^{(u)}(0)\right\rangle \\
 & =-\sum_{a}\left|\eta_{a}\right|^{2}Q_{a}\int\limits _{-\infty}^{\infty}dt\int\limits _{-\infty}^{\infty}d\tau\left\langle j^{(u)}(0)\left(A_{a}(t)A_{a}^{\dagger}(\tau)-A_{a}^{\dagger}(t)A_{a}(\tau)\right)+\left(A_{a}(t)A_{a}^{\dagger}(\tau)-A_{a}^{\dagger}(t)A_{a}(\tau)\right)j^{(u)}(0)\right\rangle
\end{align*}
where in the third line we only kept charge-conserving terms, and
in the last line we switched dummy indices for half the terms. We
can also note that by switching between $t$ and $\tau$ in the second
term, we just obtain that the second term is the complex conjugate
of the first. Separating this into products of terms given to us by
Eqs.~\eqref{eq:Current_and_vertex} and \eqref{eq:Two_Tunnel_Operator_Function},
we obtain
\begin{align}
S_{0T} & =\sum_{a}\frac{Q_{a}^{2}}{\pi}\left|\eta_{a}\right|^{2}\prod_{l}\left[\frac{1}{v_{l}^{(u)}}\frac{1}{v_{l}^{(d)}}\right]^{a_{l}^{2}}\int dt\int d\tau\frac{i\left(\pi k_{B}T_{1}\right)^{2\Delta+1}\left(\pi k_{B}T_{2}\right)^{2\Delta}\cos{\left(Q_{a}V(t-\tau)\right)}}{\left[i\sinh\left(\pi k_{B}T_{1}\left(t-\tau-i\left(\kappa-\varepsilon\right)\right)\right)i\sinh\left(\pi k_{B}T_{2}\left(t-\tau+i\left(\kappa-\varepsilon\right)\right)\right)\right]^{2\Delta}}\times\label{eq:SOT_pre_coth}\\
 & \times\left[\coth\left(\pi k_{B}T_{1}\left(-t-i\varepsilon\right)\right)-\coth\left(\pi k_{B}T_{1}\left(-\tau-i\kappa\right)\right)\right]+\text{c.c.},
\end{align}
where we employ here two positive short time cutoffs, with $\kappa>\varepsilon$.
Assuming $v_{l}^{(u)}=v_{l}^{(d)}=v_{l}$, and changing variables
to $t=\tau+y$, we obtain
\begin{align*}
S_{0T} & =\sum_{a}\frac{Q_{a}^{2}}{\pi}\left|\eta_{a}\right|^{2}\left[\prod_{l}v_{l}^{-2a_{l}^{2}}\right]\int d\tau\int dy\frac{i\left(\pi k_{B}T_{1}\right)^{2\Delta+1}\left(\pi k_{B}T_{2}\right)^{2\Delta}\cos{\left(Q_{a}Vy\right)}}{\left[i\sinh\left(\pi k_{B}T_{1}\left(y-i\left(\kappa-\varepsilon\right)\right)\right)i\sinh\left(\pi k_{B}T_{2}\left(y-i\left(\kappa-\varepsilon\right)\right)\right)\right]^{2\Delta}}\times\\
 & \times\left[\coth\left(\pi k_{B}T_{1}\left(-\tau-y-i\varepsilon\right)\right)-\coth\left(\pi k_{B}T_{1}\left(-\tau-i\kappa\right)\right)\right]+\text{c.c.},
\end{align*}
We now calculate the $\tau$ integral. The integral of a single
hyperbolic cotangent is diverging,
\[
\int\limits _{-\infty}^{\infty}d\tau\coth\left(\pi k_{B}T_{1}\left(-\tau-y-i\varepsilon\right)\right)=\int\limits _{-\infty}^{\infty}d\tau\frac{\cosh\left(\pi k_{B}T_{1}\left(-\tau-y-i\varepsilon\right)\right)}{\sinh\left(\pi k_{B}T_{1}\left(-\tau-y-i\varepsilon\right)\right)}=-\frac{1}{\pi k_{B}T_{1}}\ln\left[\sinh\left(\pi k_{B}T_{1}\left(-\tau-y-i\varepsilon\right)\right)\right]_{-\infty}^{\infty},
\]
but since we have a difference between two hyperbolic cotangents,
this divergence cancels out, and we obtain

\begin{align*}
\int\limits _{-\infty}^{\infty}d\tau\left[\coth\left(\pi k_{B}T_{1}\left(-\tau-y-i\varepsilon\right)\right)-\coth\left(\pi k_{B}T_{1}\left(-\tau-i\kappa\right)\right)\right] & =-\frac{1}{\pi k_{B}T_{1}}\ln\left[\frac{\sinh\left(\pi k_{B}T_{1}\left(-\tau-y-i\varepsilon\right)\right)}{\sinh\left(\pi k_{B}T_{1}\left(-\tau-i\kappa\right)\right)}\right]_{-\infty}^{\infty}\\
=-\frac{1}{\pi k_{B}T_{1}}\left[\ln e^{\pi k_{B}T_{1}\left(y-i\left(\kappa-\varepsilon\right)\right)}-\ln e^{-\pi k_{B}T_{1}\left(y-i\left(\kappa-\varepsilon\right)\right)}\right] & =-2\left(y-i\left(\kappa-\varepsilon\right)\right).
\end{align*}

Plugging this back into Eq.~\eqref{eq:SOT_pre_coth}, renaming $-i\left(\kappa-\varepsilon\right)\rightarrow-i\varepsilon$,
we obtain the full integral expression
\begin{equation}
S_{0T}=-4\sum_{a}\frac{Q_{a}^{2}}{\pi}\left|\eta_{a}\right|^{2}\left[\prod_{l}v_{l}^{-2a_{l}^{2}}\right]\left(\pi k_{B}T_{2}\right)^{4\Delta-1}\left(\frac{\pi k_{B}T_{1}}{\pi k_{B}T_{2}}\right)^{2\Delta-1}\int\limits _{-\infty}^{\infty}dy\frac{i\left(\pi k_{B}T_{1}\right)^{2}\left(y-i\varepsilon\right)\cos{\left(Q_{a}Vy\right)}}{\left[i\sinh\left(\pi k_{B}T_{1}\left(y-i\varepsilon\right)\right)i\sinh\left(\pi k_{B}T_{2}\left(y-i\varepsilon\right)\right)\right]^{2\Delta}},\label{eq:Full_SOT_No_Manipulations}
\end{equation}
where we replaced the $+\text{c.c.}$ with an overall factor of $2$
because we see this quantity is real.

Numerical calculation of the above integral requires treating the
vicinity of $y=0$ with care. We derive a numerics-friendly expression,
which does not require special treatment for any part of the integral,
using a convenient change of variables in the complex plane \citep{Martin2005}.
For $T_{1}\geq T_{2}$ , we define
\begin{equation}
\tau=y-i\varepsilon+\frac{i}{2k_{B}T_{1}},\label{Complex_shift}
\end{equation}
and we can write the integral as

\[
\int\limits _{-\infty-i\varepsilon+\frac{i}{2k_{B}T_{1}}}^{\infty-i\varepsilon+\frac{i}{2k_{B}T_{1}}}d\tau\frac{i\left(\pi k_{B}T_{1}\right)^{2}\left(\tau-\frac{i}{2k_{B}T_{1}}\right)\cos{\left(Q_{a}V\left(\tau-\frac{i}{2k_{B}T_{1}}\right)\right)}}{\left[i\sinh\left(\pi k_{B}T_{1}\tau-\frac{i\pi}{2}\right)i\sinh\left(\pi k_{B}T_{2}\tau-\frac{i\pi k_{B}T_{2}}{2k_{B}T_{1}}\right)\right]^{2\Delta}}.
\]
Since $i\sinh\left(\pi k_{B}T_{1}\tau-\frac{i\pi}{2}\right)=\cosh\left(\pi k_{B}T_{1}\tau\right)$,
we hence obtain
\[
\int\limits _{-\infty-i\varepsilon+\frac{i}{2k_{B}T_{1}}}^{\infty-i\varepsilon+\frac{i}{2k_{B}T_{1}}}d\tau\frac{i\left(\pi k_{B}T_{1}\right)^{2}\left(\tau-\frac{i}{2k_{B}T_{1}}\right)\cos{\left(Q_{a}V\left(\tau-\frac{i}{2k_{B}T_{1}}\right)\right)}}{\left[\cosh\left(\pi k_{B}T_{1}\tau\right)i\sinh\left(\pi k_{B}T_{2}\tau-\frac{i\pi k_{B}T_{2}}{2k_{B}T_{1}}\right)\right]^{2\Delta}}.
\]
Now defining $y=\pi k_{B}T_{1}\tau$, $\lambda=T_{1}/T_{2}$ we have
\[
\int\limits _{-\infty-i\varepsilon+\frac{i\pi}{2}}^{\infty-i\varepsilon+\frac{i\pi}{2}}dy\frac{i\left(y-\frac{i\pi}{2}\right)\cos{\left(\frac{Q_{a}V}{\pi k_{B}T_{1}}\left(y-\frac{i\pi}{2}\right)\right)}}{\left[\cosh\left(y\right)i\sinh\left(\frac{1}{\lambda}\left(y-\frac{i\pi}{2}\right)\right)\right]^{2\Delta}}.
\]

This expression has poles at $y=\frac{i\pi}{2}+i\pi n$ due to the
$\cosh$ term in the denominator, and at $y=\frac{i\pi}{2}+\lambda i\pi n$
due to the $\sinh$ term in the denominator. For $\lambda\geq1$,
we have no poles between $0$ and $\frac{i\pi}{2}-i\varepsilon$.
So we can move the contour back to the real axis, giving us a final
integral form for the noise term,
\begin{equation}
S_{0T}=-\frac{4}{\pi}\sum_{a}Q_{a}^{2}G_{a}\int\limits _{-\infty}^{\infty}dy\frac{i\left(y-\frac{i\pi}{2}\right)\cos{\left(\frac{Q_{a}V}{\pi k_{B}T_{1}}\left(y-\frac{i\pi}{2}\right)\right)}}{\left[\cosh\left(y\right)i\sinh\left(\frac{1}{\lambda}\left(y-\frac{i\pi}{2}\right)\right)\right]^{2\Delta}},\label{eq:SOT_Integral_Form}
\end{equation}
where $G_{a}$ is given in Eq.~(\ref{eq:Effective_Tunneling_Constant}).

This is a convenient expression, which can be calculated numerically
without any special care. Finally, we note that for $S_{0T}^{(d)}$,
the entire derivation should be the same, except we lose one factor
of $\lambda$ due to replacing $T_{1}\leftrightarrow T_{2}$ in Eq.~(\ref{eq:SOT_pre_coth}).
In the case of $T_{2}>T_{1}$, the derivation is also rather similar,
with the only difference being that the shift in the complex plane described
in Eq.~\eqref{Complex_shift} is now $\tau=y-i\varepsilon+\frac{i}{2k_{B}T_{2}}.$

\subsubsection{$S_{TT}$}

Plugging Eq.~\eqref{eq:Tunneling_current} into Eq.~\eqref{eq:Define_noise_terms},
and only keeping charge-conserving terms, we obtain
\begin{equation}
S_{TT}=\sum_{a}Q_{a}^{2}\left|\eta_{a}\right|^{2}\int dt\left\langle A_{a}(0)A_{a}^{\dagger}(t)+A_{a}^{\dagger}(0)A_{a}(t)+A_{a}(t)A_{a}^{\dagger}(0)+A_{a}^{\dagger}(t)A_{a}(0)\right\rangle .
\end{equation}
Plugging in the values found above for all correlation functions,
assuming that $v_{l}^{(u)}=v_{l}^{(d)}=v_{l}$, and that we have a
scaling dimension of $\Delta\equiv\sum_{l}\frac{a_{l}^{2}}{2}$, this
now gives
\begin{equation}
S_{TT}=4\sum_{a}Q_{a}^{2}\left|\eta_{a}\right|^{2}\left[\prod_{l}v_{l}^{-2a_{l}^{2}}\right]\int\limits _{-\infty}^{\infty}dt\frac{\left(\pi k_{B}T_{1}\right)^{2\Delta}\left(\pi k_{B}T_{2}\right)^{2\Delta}\cos{\left(Q_{a}Vt\right)}}{\left[i\sinh\left(\pi k_{B}T_{1}\left(t-i\varepsilon\right)\right)i\sinh\left(\pi k_{B}T_{2}\left(t-i\varepsilon\right)\right)\right]^{2\Delta}}.\label{eq:Full_STT_No_Manipulations}
\end{equation}

We continue with the same manipulations used to convert Eq.~\eqref{eq:Full_SOT_No_Manipulations}
to Eq.~\eqref{eq:SOT_Integral_Form}, using a change of variables,
$\tau=t-i\varepsilon+\frac{i}{2k_{B}\text{max}(T_{1},T_{2})}$ and
utilizing the location of the poles in the denominator to shift our
contour back to the real axis. This gives us the convenient integral
expression for $\lambda\geq1$
\begin{equation}
S_{TT}=4\sum_{a}Q_{a}^{2}G_{a}\int\limits _{-\infty}^{\infty}dy\frac{\cos{\left(\frac{Q_{a}V}{\pi k_{B}T_{1}}\left(y-\frac{i\pi}{2}\right)\right)}}{\left[\cosh\left(y\right)i\sinh\left(\frac{1}{\lambda}\left(y-\frac{i\pi}{2}\right)\right)\right]^{2\Delta}},\label{eq:STT_Integral_Form}
\end{equation}
with the extension to the case $\lambda<1$ being straight-forward.

\subsection{Current}

Calculation of the average tunneling current is very similar to the
calculations above. Going up an order in the Kubo formula, we obtain

\[
\langle\hat{I}_{T}\rangle=-i\int\limits _{-\infty}^{t}d\tau\left\langle \left[\hat{I}_{T}(t),\sum_{a}\left(\eta_{a}A_{a}(\tau)+\eta_{a}^{*}A_{a}^{\dagger}(\tau)\right)\right]\right\rangle .
\]
The appropriate utilization of Eqs.~\eqref{eq:Tunneling_current}
and \eqref{eq:Two_Tunnel_Operator_Function} lead to

\begin{equation}
I_{T}=2i\sum_{a}Q_{a}\left|\eta_{a}\right|^{2}\left[\prod_{l}v_{l}^{-2a_{l}^{2}}\right]\int\limits _{-\infty}^{\infty}dt\frac{\left(\pi k_{B}T_{1}\right)^{2\Delta}\left(\pi k_{B}T_{2}\right)^{2\Delta}\sin\left(Q_{a}Vt\right)}{\left[i\sinh\left(\pi k_{B}T_{1}\left(t-i\varepsilon\right)\right)i\sinh\left(\pi k_{B}T_{2}\left(t-i\varepsilon\right)\right)\right]^{2\Delta}},\label{eq:Full_Current_No_Manipulations}
\end{equation}
where we define $I_{T}\equiv\left\langle \hat{I}_{T}\right\rangle $.
Similar manipulations as the two previous sections lead to the final
expression for $\lambda\geq1$
\begin{equation}
I_{T}=2i\sum_{a}Q_{a}G_{a}\int\limits _{-\infty}^{\infty}dy\frac{\sin\left(\frac{Q_{a}V}{\pi k_{B}T_{1}}\left(y-\frac{i\pi}{2}\right)\right)}{\left[\cosh\left(y\right)i\sinh\left(\frac{1}{\lambda}\left(y-\frac{i\pi}{2}\right)\right)\right]^{2\Delta}},\label{eq:Current_Integral_Form}
\end{equation}
which is numerically convergent, and predictably yields finite current
only for finite voltage.

\subsection{Limits}

\label{app:Limits} Here we show how the expressions in Eqs.~(\ref{eq:SOT_Integral_Form}),
(\ref{eq:STT_Integral_Form}), and~(\ref{eq:Current_Integral_Form})
reduce to more convenient expressions in the regimes discussed in
Section~\ref{sec:IV_experimentally_relevant_scenarios}. The extension
to the case $\lambda<1$ is again straight-forward.

\subsubsection{Equal temperatures}

\label{app:Equal_Temperatures} For equal temperatures, we define
$T_{1}=T_{2}\equiv T$. As such, in all three integral expressions,
we may replace $i\sinh\left(\frac{1}{\lambda}\left(y-\frac{i\pi}{2}\right)\right)\rightarrow\cosh\left(y\right)$.
The three terms now give

\begin{align}
S_{0T} & =-\frac{4}{\pi}\sum_{a}Q_{a}^{2}G_{a}\int\limits _{-\infty}^{\infty}dy\frac{i\left(y-\frac{i\pi}{2}\right)\cos{\left(\frac{Q_{a}V}{\pi k_{B}T}\left(y-\frac{i\pi}{2}\right)\right)}}{\left[\cosh\left(y\right)\right]^{4\Delta}},\\
S_{TT} & =4\sum_{a}Q_{a}^{2}G_{a}\int\limits _{-\infty}^{\infty}dy\frac{\cos{\left(\frac{Q_{a}V}{\pi k_{B}T}\left(y-\frac{i\pi}{2}\right)\right)}}{\left[\cosh\left(y\right)\right]^{4\Delta}},\\
\left\langle \hat{I}_{T}\right\rangle  & =2i\sum_{a}Q_{a}G_{a}\int\limits _{-\infty}^{\infty}dy\frac{\sin\left(\frac{Q_{a}V}{\pi k_{B}T}\left(y-\frac{i\pi}{2}\right)\right)}{\left[\cosh\left(y\right)\right]^{4\Delta}}.
\end{align}
and the excess noise is given by
\begin{equation}
S=S_{TT}+2S_{0T}=-i\frac{8}{\pi}\sum_{a}Q_{a}^{2}G_{a}\int\limits _{-\infty}^{\infty}dy\frac{y\cos{\left(\frac{Q_{a}V}{\pi k_{B}T}\left(y-\frac{i\pi}{2}\right)\right)}}{\left[\cosh\left(y\right)\right]^{4\Delta}}.
\end{equation}
Since all denominators are now even, we may keep only the even components
of the respective numerators, reducing the expressions to

\begin{align}
S & =\frac{8}{\pi}\sum_{a}Q_{a}^{2}G_{a}\sinh\left(\frac{Q_{a}V}{2k_{B}T_{0}}\right)\int\limits _{-\infty}^{\infty}dy\frac{y\sin{\left(\frac{Q_{a}V}{\pi k_{B}T}y\right)}}{\left[\cosh\left(y\right)\right]^{4\Delta}},\\
\left\langle \hat{I}_{T}\right\rangle  & =2\sum_{a}Q_{a}G_{a}\sinh\left(\frac{Q_{a}V}{2k_{B}T_{0}}\right)\int\limits _{-\infty}^{\infty}dy\frac{\cos\left(\frac{Q_{a}V}{\pi k_{B}T}y\right)}{\left[\cosh\left(y\right)\right]^{4\Delta}}.
\end{align}
These are now well-known integrals, which correspond to

\begin{align}
\int\limits _{-\infty}^{\infty}dy\frac{\cos\left(\frac{Q_{a}V}{\pi k_{B}T}y\right)}{\left[\cosh\left(y\right)\right]^{4\Delta}} & =2^{4\Delta-1}\mathcal{B}\left(2\Delta+i\frac{Q_{a}V}{2\pi k_{B}T},2\Delta-i\frac{Q_{a}V}{2\pi k_{B}T}\right),\\
\int\limits _{-\infty}^{\infty}dy\frac{y\sin{\left(\frac{Q_{a}V}{\pi k_{B}T}y\right)}}{\left[\cosh\left(y\right)\right]^{4\Delta}} & =-\frac{\pi k_{B}T}{Q_{a}}\frac{\partial}{\partial V}\left[\int\limits _{-\infty}^{\infty}dy\frac{\cos\left(\frac{Q_{a}V}{\pi k_{B}T}y\right)}{\left[\cosh\left(y\right)\right]^{4\Delta}}\right]\\
 & =2^{4\Delta-1}\mathcal{B}\left(2\Delta+i\frac{Q_{a}V}{2\pi k_{B}T},2\Delta-i\frac{Q_{a}V}{2\pi k_{B}T}\right)\mathrm{Im}\left[\psi\left(2\Delta+i\frac{Q_{a}V}{2\pi k_{B}T}\right)\right],
\end{align}
where $\mathcal{B}(x,y)$ is the beta function, the digamma function $\psi(x)=\Gamma'(x)/\Gamma(x)$
is the logarithmic derivative of the gamma function, and $\mathrm{Im}$
stands for the imaginary part. Finally, for a single vector $a,Q_{a}\equiv Q$,
this reduces to

\begin{equation}
F\equiv\frac{S}{2eI_{T}}=\frac{2Q}{\pi e}\mathrm{Im}\left[\psi\left(2\Delta+i\frac{QV}{2\pi k_{B}T}\right)\right].
\end{equation}

\subsubsection{Dominant Temperature}

\label{app:Dominant_Temperature}

In the regime where bias voltage is much smaller than the two temperatures,
i.e., $eV\ll k_{B}T_{1},k_{B}T_{2}$, we may expand all trigonometric
functions to the leading order in $\frac{eV}{k_{B}T_{i}}$. The three
terms now give

\begin{align}
S_{0T} & =-\frac{4}{\pi}\sum_{a}Q_{a}^{2}G_{a}\int\limits _{-\infty}^{\infty}dy\frac{i\left(y-\frac{i\pi}{2}\right)}{\left[\cosh\left(y\right)i\sinh\left(\frac{1}{\lambda}\left(y-\frac{i\pi}{2}\right)\right)\right]^{2\Delta}},\label{eq:Full_Integral_Expressions_Dominant_Temp}\\
S_{TT} & =4\sum_{a}Q_{a}^{2}G_{a}\int\limits _{-\infty}^{\infty}dy\frac{1}{\left[\cosh\left(y\right)i\sinh\left(\frac{1}{\lambda}\left(y-\frac{i\pi}{2}\right)\right)\right]^{2\Delta}},\\
I_{T} & =2i\sum_{a}Q_{a}G_{a}\int\limits _{-\infty}^{\infty}dy\frac{\frac{Q_{a}V}{\pi k_{B}T_{1}}\left(y-\frac{i\pi}{2}\right)}{\left[\cosh\left(y\right)i\sinh\left(\frac{1}{\lambda}\left(y-\frac{i\pi}{2}\right)\right)\right]^{2\Delta}}.
\end{align}

We note that the integrals for $S_{0T}$ and $I_{T}$ are now completely
equivalent. Assuming only one type of quasiparticle tunnels, the Fano
factor is now given by

\begin{equation}
F=\frac{S_{TT}+2S_{0T}}{2eI_{T}}=\frac{\pi k_{B}T_{1}}{eV}\left[\frac{\int\limits _{-\infty}^{\infty}dy\frac{1}{\left[\cosh\left(y\right)i\sinh\left(\frac{1}{\lambda}\left(y-\frac{i\pi}{2}\right)\right)\right]^{2\Delta}}}{\int\limits _{-\infty}^{\infty}dy\frac{i\left(y-\frac{i\pi}{2}\right)}{\left[\cosh\left(y\right)i\sinh\left(\frac{1}{\lambda}\left(y-\frac{i\pi}{2}\right)\right)\right]^{2\Delta}}}-\frac{2}{\pi}\right].
\end{equation}

In the limit where $\lambda\gg1$, we can further simplify this expression
by approximating $\sinh\left(\frac{1}{\lambda}\left(y-\frac{i\pi}{2}\right)\right)\approx\frac{1}{\lambda}\left(y-\frac{i\pi}{2}\right)$,
such that the Fano factor is now to the leading order

\begin{equation}
F=\frac{k_{B}T_{1}}{eV}\left[\frac{\pi\int\limits _{-\infty}^{\infty}dy\frac{1}{\left[\cosh\left(y\right)i\left(y-\frac{i\pi}{2}\right)\right]^{2\Delta}}}{\int\limits _{-\infty}^{\infty}dy\frac{i\left(y-\frac{i\pi}{2}\right)}{\left[\cosh\left(y\right)i\left(y-\frac{i\pi}{2}\right)\right]^{2\Delta}}}-2\right]\equiv f(\Delta)\frac{k_{B}T_{1}}{eV}.
\end{equation}
This defines the function $f(\Delta)$ used in Eq.~\eqref{eq:Fano_diffTemps_highT1_asymp}.

\subsubsection{Dominant voltage}

\label{app:Dominant Voltage}

In the case of dominant voltage, $eV\gg k_{B}T_{i}$ we return to
Eqs. \eqref{eq:Full_SOT_No_Manipulations}, \eqref{eq:Full_STT_No_Manipulations}
and \eqref{eq:Full_Current_No_Manipulations}, and redefine $y=Q_{a}Vt$.
This gives us the expressions

\begin{align}
S_{0T} & =-\frac{4}{\pi}\sum_{a}Q_{a}^{2}\left|\eta_{a}\right|^{2}\left[\prod_{l}v_{l}^{-2a_{l}^{2}}\right]\int\limits _{-\infty}^{\infty}\frac{dy}{(Q_{a}V)^{2}}\frac{i\left(\pi k_{B}T_{1}\right)^{2\Delta+1}\left(\pi k_{B}T_{2}\right)^{2\Delta}\left(y-i\varepsilon\right)\cos{\left(y\right)}}{\left[i\sinh\left(\frac{\pi k_{B}T_{1}}{Q_{a}V}\left(y-i\varepsilon\right)\right)i\sinh\left(\frac{\pi k_{B}T_{2}}{Q_{a}V}\left(y-i\varepsilon\right)\right)\right]^{2\Delta}},\\
S_{TT} & =4\sum_{a}Q_{a}^{2}\left|\eta_{a}\right|^{2}\left[\prod_{l}v_{l}^{-2a_{l}^{2}}\right]\int\limits _{-\infty}^{\infty}\frac{dy}{Q_{a}V}\frac{\left(\pi k_{B}T_{1}\right)^{2\Delta}\left(\pi k_{B}T_{2}\right)^{2\Delta}\cos{\left(y\right)}}{\left[i\sinh\left(\frac{\pi k_{B}T_{1}}{Q_{a}V}\left(y-i\varepsilon\right)\right)i\sinh\left(\frac{\pi k_{B}T_{2}}{Q_{a}V}\left(y-i\varepsilon\right)\right)\right]^{2\Delta}},\\
I_{T} & =2i\sum_{a}Q_{a}\left|\eta_{a}\right|^{2}\left[\prod_{l}v_{l}^{-2a_{l}^{2}}\right]\int\limits _{-\infty}^{\infty}\frac{dy}{Q_{a}V}\frac{\left(\pi k_{B}T_{1}\right)^{2\Delta}\left(\pi k_{B}T_{2}\right)^{2\Delta}\sin\left(y\right)}{\left[i\sinh\left(\frac{\pi k_{B}T_{1}}{Q_{a}V}\left(y-i\varepsilon\right)\right)i\sinh\left(\frac{\pi k_{B}T_{2}}{Q_{a}V}\left(y-i\varepsilon\right)\right)\right]^{2\Delta}}.
\end{align}

Now replacing $\sinh\left(\frac{\pi k_{B}T_{i}}{Q_{a}V}\left(y-i\varepsilon\right)\right)\approx\frac{\pi k_{B}T_{i}}{Q_{a}V}\left(y-i\varepsilon\right)$,
these expressions now give to the leading order

\begin{align}
S_{0T} & =-4\sum_{a}\frac{Q_{a}^{2}}{\pi}\left|\eta_{a}\right|^{2}\left[\prod_{l}v_{l}^{-2a_{l}^{2}}\right]\frac{\pi k_{B}T_{1}}{Q_{a}V}\int\limits _{-\infty}^{\infty}\frac{dy}{Q_{a}V}\frac{\left(Q_{a}V\right)^{4\Delta}\cos{\left(y\right)}}{\left[i\left(y-i\varepsilon\right)\right]^{4\Delta-1}},\\
S_{TT} & =4\sum_{a}Q_{a}^{2}\left|\eta_{a}\right|^{2}\left[\prod_{l}v_{l}^{-2a_{l}^{2}}\right]\int\limits _{-\infty}^{\infty}\frac{dy}{Q_{a}V}\frac{\left(Q_{a}V\right)^{4\Delta}\cos{\left(y\right)}}{\left[i\left(y-i\varepsilon\right)\right]^{4\Delta}},\\
I_{T} & =2i\sum_{a}Q_{a}\left|\eta_{a}\right|^{2}\left[\prod_{l}v_{l}^{-2a_{l}^{2}}\right]\int\limits _{-\infty}^{\infty}\frac{dy}{Q_{a}V}\frac{\left(Q_{a}V\right)^{4\Delta}\sin\left(y\right)}{\left[i\left(y-i\varepsilon\right)\right]^{4\Delta}}.
\end{align}

These integrals are now known in terms of Euler gamma functions. For
a single quasiparticle, this gives the Fano factor of

\begin{equation}
F=\frac{S_{TT}+2S_{0T}}{2eI_{T}}=\frac{Q}{e}+\frac{k_{B}T_{1}}{eV}\left(1-4\Delta\right)+O\left[\left(\frac{k_{B}T_{i}}{eV}\right)^{2}\right].
\end{equation}
\end{widetext}

\end{document}